\documentclass[aip, reprint, floatfix]{revtex4-1}

\usepackage{bm}
\usepackage{amsmath}
\usepackage{graphicx}
\usepackage{float}

\DeclareMathOperator{\erf}{erf}

\usepackage{tikz}
\usepackage{pgfplots}
\pgfplotsset{compat=newest,/pgf/number format/.cd,1000 sep={}}
\pgfplotsset{
  layers/axis lines on top/.define layer set={
    axis background,
    axis grid,
    axis tick labels,
    pre main,
    main,
    axis ticks,
    axis lines,
    axis descriptions,
    axis foreground,
  }{/pgfplots/layers/standard},
}
\usepackage{pgfplotstable}

\usetikzlibrary{external}
\usetikzlibrary{arrows.meta}

\usepgfplotslibrary{colormaps} 
\usepgfplotslibrary{polar}
\pgfplotsset{colormap={inferno}{
    rgb=(1.0,1.0,1.0)  
    rgb=(0.99,0.88,0.55) 
    rgb=(0.99,0.80,0.52) 
    rgb=(0.98,0.75,0.52) 
    rgb=(0.96,0.63,0.43) 
    rgb=(0.94,0.50,0.37) 
    rgb=(0.90,0.35,0.30) 
    rgb=(0.85,0.30,0.30) 
    rgb=(0.75,0.21,0.36) 
    rgb=(0.65,0.18,0.38) 
    rgb=(0.62,0.14,0.39) 
    rgb=(0.49,0.12,0.38) 
    rgb=(0.35,0.11,0.33) 
    rgb=(0.28,0.11,0.32) 
    rgb=(0.23,0.10,0.30) 
    rgb=(0.18,0.09,0.29) 
}}

\usepackage{subcaption}

\definecolor{General}{RGB}{34, 42, 53}
\definecolor{uniSblue}{RGB}{0,81,158}
\definecolor{DarkGreen}{RGB}{51,160,44}
\definecolor{DarkRed}{RGB}{186, 74, 50}
\definecolor{customblue}{RGB}{55,126,184}
\definecolor{customred}{RGB}{228,26,28}

\usepackage{makecell}

\begin{document}


\title[Machine Learning-Based Scattering Kernel for Very Low Earth Orbit]{A Machine Learning Framework for Scattering Kernel Derivation Using Molecular Dynamics Data in Very Low Earth Orbit} 

\author{M. Schütte}
\email{schuettem@irs.uni-stuttgart.de}
\affiliation{Institute of Space Systems, University of Stuttgart}%
\homepage{https://www.irs.uni-stuttgart.de/en/}

\author{S. Hocker}%
\email{stephen.hocker@fmq.uni-stuttgart.de}
\affiliation{Institute for Functional Matter and Quantum Technologies, University of Stuttgart}
\homepage{https://www.fmq.uni-stuttgart.de/}

\author{H. Lipp}
\affiliation{Institute for Functional Matter and Quantum Technologies, University of Stuttgart}

\author{J. Roth}
\affiliation{Institute for Functional Matter and Quantum Technologies, University of Stuttgart}

\author{S. Fasoulas}
\affiliation{Institute of Space Systems, University of Stuttgart}

\author{M. Pfeiffer}
\affiliation{Institute of Space Systems, University of Stuttgart}

\date{\today, Preprint Submitted to Physics of Fluids}

\begin{abstract}

    The free molecular flow regime in VLEO makes gas-surface interactions (GSIs) crucial for satellite aerodynamic modeling.
    The Direct Simulation Monte Carlo (DSMC) method is required to estimate aerodynamic forces due to the breakdown of the continuum assumption.
    In DSMC, the Maxwell model is the most widely used approach for GSI.
    It simplifies the process by treating it as a superposition of diffuse and specular reflections while assuming a constant accommodation coefficient.
    In reality, this coefficient is influenced by multiple factors, such as the angle and magnitude of the incident velocity.
    A high-precision GSI model could significantly improve satellite aerodynamics optimization and the design of efficient intakes for atmospheric breathing propulsion systems.
    This advancement would greatly refine mission planning and fuel requirement calculations, ultimately extending operational lifetimes and lowering costs.
    To gain a deep understanding of the GSI at the microscopic level, molecular dynamics (MD) simulations provide valuable insights into the physical processes involved.
    However, due to computational limitations, simulating an entire satellite is impractical.
    Instead, we use MD to analyze the impact of selected velocity vectors on a amorphous $\text{Al}_2\text{O}_3$ surface.
    The obtained scattering kernels for the respective velocity vectors are then used to train a conditional Variational Autoencoder (cVAE).
    This model is able to generate scattering kernels for any incident velocity vector and can be integrated into DSMC simulations, significantly enhancing their accuracy.
    Applications of this model on a flat plate have shown that the cVAE is able to predict the shift from diffuse to quasi-specular reflection with increasing polar angle.
    Additionally, the aerodynamic coefficients and molecular fluxes are considerably different from those obtained with the Maxwell model.

\end{abstract}


\maketitle 

\section{Introduction} \label{sec:introduction}

The Very Low Earth Orbit (VLEO), typically defined as orbital altitudes below $450\,$km, has gained significant attention within the aerospace sector due to its unique advantages.
Among its benefits are reduced launch costs, improved measurement accuracy, and a decreased risk of collisions with space debris, as a result of the self-cleaning effect caused by the residual atmosphere \cite{hao_very_2014,crisp_benefits_2020}.
Nevertheless, the residual atmosphere in VLEO presents challenges for satellite operations.
Satellites encounter an increased atmospheric drag, which decreases their operational lifespan and necessitates the use of propulsion systems to counteract this drag.
However, it is also possible to utilize the residual atmosphere to the advantage of the satellite.
Atomic oxygen, which is the primary component in these altitudes, can be used as a propellant for atmospheric breathing electric propulsion systems (ABEPs).
The resulting aerodynamic forces can also be harnessed to control the altitude and orbit of the satellite, reducing the need for thrusters and fuel \cite{hao_very_2014,crisp_benefits_2020}.
The collaborative research center ATLAS (Advancing Technologies for Low-Altitude Satellites) is dedicated to overcoming the key challenges associated with making VLEO accessible \cite{fasoulas_atlas_2025}.
One of the primary focuses within ATLAS is the investigation of gas-surface interaction (GSI) and the development of an advanced gas-surface scattering model for Direct Simulation Monte Carlo (DSMC) simulations.
This model is not limited to DSMC applications but can also be employed in the Test Particle Monte Carlo (TPMC) method.
Furthermore, the aerodynamic coefficients derived from this model can be incorporated into panel methods, significantly improving aerodynamic predictions.

The DSMC method is suitable for the accurate modeling of non-equilibrium gas flows, such as those encountered in the VLEO regime \cite{bird_molecular_1994}.
For optimization of satellite aerodynamics and the development of efficient ABEP intakes, it is crucial to precisely model the GSIs.
The most common GSI model used in DSMC simulations is the Maxwell model, which was first introduced by James Clerk Maxwell in 1879 \cite{maxwell_stresses_1879}.
It models the interaction either as a specular reflection, where the incident particle is reflected with the same velocity but in the opposite direction, or as a diffuse reflection, where the incident particle is scattered according to a cosine distribution of the molecular flux.
However, GSIs are much more complex.
Molecular beam experiments have shown that the reflected particles exhibit a lobular distribution of the molecular flux \cite{xu_hyperthermal_2025, minton,murray_experiment_2017, poovathingal_experiments_2016}, which is not captured by the Maxwell model.
Consequently, a variety of models have been developed, such as the Cercignani-Lampis model \cite{cercignani_kinetic_1971} and the Washboard model \cite{tully_washboard_1990}.
All of these models use simplifications, such as independent scattering in the normal and tangential directions.
Furthermore, they depend on accommodation coefficients or other constant parameters.
These parameters are usually unknown and depend on various factors such as gas species, surface material, surface temperature, gas temperature, gas velocity, angle of incidence, absorbed atomic oxygen and surface roughness \cite{liang_parameter-free_2021,mehta_comparing_2014}.

In contrast to traditional scattering models, there exists a relatively new category of data-driven scattering kernels.
Such models use molecular dynamics (MD) simulations to get detailed insights into the microscopic behavior of GSIs.
For data generation, they simulated the GSIs of thousands of different impinging velocity vectors once.
As a result, they derive the joint probability density function $P(\bm{v}_{\text{i}}, \bm{v}_{\text{r}})$.
Afterwards, the joint probability density function is described using a Gaussian mixture model (GMM) \cite{liao_prediction_2018,nejad_scattering_2023} or a distribution element tree (DET) method \cite{andric_det_2019}.
From the resulting methods, it is possible to derive the scattering kernel for a given impinging velocity.
However, these methods are extremely limited to the simulated impact velocity range.
If a particle has a velocity outside this range, the joint probability function of the GMM and DET approaches zero.
Consequently, extrapolation is not possible for these methods.
Furthermore, these methods were set up for easy test cases like a Fourier flow and Couette flow without a bulk velocity.
As a result, only the gas temperature determines the incident velocity vectors.
Even for this limited range, it was suggested that the GMM should use $100\,000$ interaction pairs $(\bm{v}_{\text{i}}, \bm{v}_{\text{r}})$ to obtain a good approximation of the joint distribution \cite{nejad_scattering_2023}.
On the other hand, the DET method suffers from statistical noise if the number of MD data is limited, and simplifications are necessary to counteract \cite{andric_det_2019}.
In the VLEO regime, the bulk velocity is not at rest.
Additionally, the bulk velocity may exhibit varying polar angles in relation to the normal direction of the satellite's local surface elements.
Thus, it would be necessary to simulate $100\,000$ interaction pairs for different incident bulk velocity vectors with regard to the surface normal.

The aim of this study is to develop an efficient data-driven scattering model for VLEO satellites using MD simulation for DSMC simulations.
Instead of simulating a large number of different particle interactions with MD, it would be more efficient to specifically select incident particle velocities.
These velocities are then interacting with the surface several times in order to derive the scattering kernel for that specific velocity.
The scattering kernels are then used to train a generative machine learning model which can predict the scattering kernel for different impinging particle velocities.
As previously mentioned, the model is also applicable to TPMC, and the aerodynamic coefficients obtained from this model can be integrated into panel methods like ADBSat \cite{sinpetru_adbsat_2022}.

\section{Methods} \label{sec:methods}

\subsection{Computational Methods} \label{subsec:computational_methods}

\subsubsection{Direct Simulation Monte Carlo} \label{subsubsec:dsmc}

The high Knudsen number in VLEO environments causes the continuum assumption to break down.
To capture the non-equilibrium behavior of the gas, simulation techniques such as the DSMC method are employed.
DSMC models gas flow by tracking the paths of individual gas particles.
Its purpose is to approximate the Boltzmann equation, which governs the time evolution of the particle distribution function $f(\bm{r}, \bm{v}, t)$ in phase space.
This function describes the probability density of finding a particle at a specific position $\bm{r}$, with velocity $\bm{v}$, at time $t$.
The Boltzmann equation is given by
\begin{equation} \label{eq:boltzmann}
    \frac{\partial f}{\partial t} + \bm{v} \cdot \frac{\partial f}{\partial \bm{r}} + \frac{\bm{F}}{m} \cdot \frac{\partial f}{\partial \bm{v}} = \left( \frac{\delta f}{\delta t} \right)_{\text{coll}}.
\end{equation}
The first term represents the time change of the distribution function.
The second term accounts for the movement of particles, which alters their spatial position $\bm{r}$ based on their velocity $\bm{v}$. 
The third term describes how external forces $\bm{F}$ influence particle velocity $\bm{v}$ by exerting acceleration proportional to the particle mass $m$. 
Finally, the right-hand side of the equation encapsulates the effects of particle collisions, which redistribute velocities and drive the system toward equilibrium.

In DSMC, this distribution function is approximated by a set of simulation particles.
Each of these simulation particles represents a weighted number of real gas molecules or atoms.
During each time step, these simulation particles move through the computational domain, interacting with each other and with boundaries, like the surfaces of the satellite.
These interactions are modeled probabilistically.
Importantly, electromagnetic interactions are typically neglected within the DSMC framework \cite{bird_molecular_1994, nejad_scattering_2023}.

In the DSMC approach, the GSIs are modeled using scattering kernels.
These kernels connect the incident velocity distribution $f(\bm{v}_{\text{i}})$ and the reflected velocity distribution $f(\bm{v}_{\text{r}})$ at the surface.
This relationship is shown in equation \eqref{eq:scattering_kernel} \cite{nejad_scattering_2023}:
\begin{equation}
    v_{\text{n,r}} \, f\left( \bm{v}_{\text{r}} \right) = \int_{v_{\text{n,i}}<0} |v_{\text{n,i}}| \, \mathcal{K}(\bm{v}_{\text{i}} \rightarrow \bm{v}_{\text{r}}) \, f\left( \bm{v}_{\text{i}} \right) \, \mathrm{d} \bm{v}_{\text{i}}
    \label{eq:scattering_kernel}
\end{equation}
where $v_{\text{n,r}}$ and $v_{\text{n,i}}$ denote the normal components of the reflected and incident velocity vectors, respectively.
Here, the scattering kernel $\mathcal{K}(\bm{v}_{\text{i}} \rightarrow \bm{v}_{\text{r}})$ acts as a conditional probability density function, describing the probability that a particle with incident velocity $\bm{v}_{\text{i}}$ will be scattered into velocity $\bm{v}_{\text{r}}$.

The Maxwell scattering kernel is perhaps the most widely used model for describing gas-surface interactions in DSMC \cite{maxwell_stresses_1879}.
It assumes that an impinging particle may be reflected either specularly or diffusely.
This model is mathematically captured by
\begin{equation} \label{eq:maxwell_kernel}
    \mathcal{K}(\bm{v}_{\text{i}} \rightarrow \bm{v}_{\text{r}}) = \underbrace{(1 - \sigma) \delta(\bm{v}_{\text{r}} - \bm{v}_{\text{i}})}_{\text{specular}} + \underbrace{\sigma f_{\text{MB}}(\bm{v}_{\text{r}}, T_{\text{w}}) |v_{\text{n,r}}|}_{\text{diffuse}}
\end{equation}
where the specular term represents mirror-like reflection, and the diffuse term describes the re-emission of particles following a Maxwell–Boltzmann distribution $f_{\text{MB}}$ at the wall temperature $T_{\text{w}}$.
In this formulation, $\delta$ is the Dirac delta function and $\sigma$ is the accommodation coefficient that quantifies the relative contributions of specular and diffuse reflections: $\sigma = 1$ implies purely diffuse behavior, while $\sigma = 0$ indicates purely specular reflection \cite{maxwell_stresses_1879, livadiotti_review_2020}.
It is commonly assumed that the accommodation coefficient remains constant for all particles and surface types.
However, this coefficient is affected by numerous factors such as the velocity of the particle, the surface temperature, the surface material, and the angle of incidence.
These dependencies are not well understood and the accommodation coefficient is usually unknown \cite{liang_parameter-free_2021,mehta_comparing_2014}.
Furthermore, molecular beam experiments have demonstrated that gas-surface interactions are not simply a linear combination of specular and diffuse reflections \cite{xu_hyperthermal_2025,murray_experiment_2017, minton,poovathingal_experiments_2016}.
As a consequence, various scattering kernels have been developed, including the Cercignani-Lampis model \cite{cercignani_kinetic_1971} and the Washboard model \cite{tully_washboard_1990}.
However, these models also rely heavily on uncertain accommodation coefficients and simplifying assumptions.
This inherent complexity makes it extremely challenging to optimize the aerodynamics of satellites in VLEO or to design efficient inlet geometries for atmospheric electric propulsion systems.
Therefore, a more accurate GSI model is needed for DSMC simulations in VLEO.

\subsubsection{Molecular Dynamics Simulations} \label{subsubsec:md}

MD simulations offer a deterministic, atomistic perspective on GSIs.
In MD, the trajectories of individual gas particles as well as surface atoms are computed.
The heart of MD lies in the calculation of interatomic forces derived from a potential energy function, $U$.
These potentials capture forces such as attraction, repulsion, and bonding between atoms, allowing for the prediction of system behavior at the atomic level.
There are several approaches to determine the potential energy function $U$.
Among them, ab initio methods, such as those based on density functional theory (DFT), provide highly accurate descriptions by solving the Schrödinger equation to determine electron distributions and forces between atoms. 
These methods are computationally intensive but are widely used for systems where high precision is necessary.
On the other hand, parameterized empirical potentials offer a simplified and computationally efficient alternative. 
These potentials are based on predetermined functional forms and parameters fitted to experimental data or ab initio data \cite{allen_md_2017,chen_molecular_2023,nejad_scattering_2023}.

After the calculation of the potential, the force acting on a given particle $j$ by all other particles in the system can be determined by the gradient of the potential energy with respect to its position:
\begin{equation}
    \bm{F}_{j} = - \nabla_{j} U(\bm{r}_1, \bm{r}_2, \dots, \bm{r}_N),
\end{equation}
where $\bm{r}_1, \bm{r}_2, \dots, \bm{r}_N$ denote the positions of all $N$ particles in the system.
These forces are then used in Newton's second law,
\begin{equation}
    \label{eq:newton}
    m_{j} \frac{\text{d}^2 \mathbf{r}_j}{\text{d}t^2} = \bm{F}_{j},
\end{equation}
which governs the time evolution of each particle's position $\bm{r}_j$ \cite{allen_md_2017,chen_molecular_2023,nejad_scattering_2023}.

Although MD simulations provide detailed insights into the atomic-level interactions between gas particles and surfaces, they are computationally intensive and typically limited to small systems and short timescales.
However, the detailed information obtained from MD simulations can be invaluable for developing more accurate scattering kernels for DSMC simulations.
Previously mentioned methods, such as the GMM or the DET method, were successfully applied to MD data to derive scattering kernels for simple cases \cite{liao_prediction_2018,nejad_scattering_2023,andric_det_2019}.
As already mentioned, to construct such a model, a large number of different incident velocity vectors would be required in the VLEO regime, because these methods are based on the joint distribution $P(\bm{v_{\text{i}}}, \bm{v_{\text{r}}})$.
It is more efficient to simulate a specific set of incident particles $\bm{v}_{\text{i}}$ with MD simulations and to derive the scattering kernel for these specific velocities.
The scattering kernels can then be used to train a generative machine learning model, such as a conditional Variational Autoencoder (cVAE), to predict the scattering kernel for any incident particle velocity.

\subsection{Data-Based Scattering Model} \label{subsec:scattering_model}

To create a data-based scattering model, we first have to generate the data using MD simulations, and then we create afterwards a cVAE on the generated data.
In the following two Sections, the data generation and the cVAE are described in detail.

\begin{figure}[tb]
    \centering
  \tikzsetnextfilename{cos}%
  \input{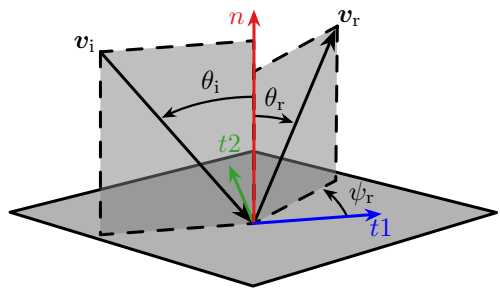}%

    \caption{Wall frame of reference ($t1$, $t2$, $n$).}
    \label{fig:cos}
\end{figure}

\begin{figure}[tb]
    \centering
  \tikzsetnextfilename{v_mag_distribution}%
  \input{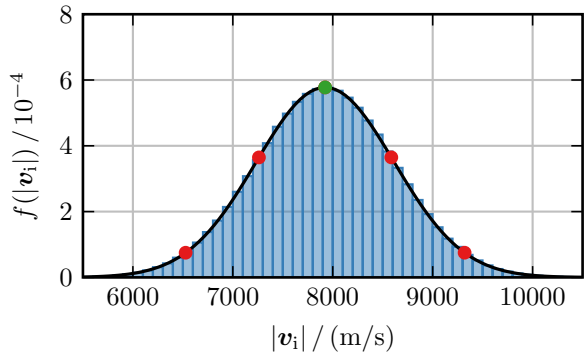}%

    \caption{Distribution of incident velocity magnitudes at an orbit velocity of $7800\,\text{m/s}$ and a gas temperature of $934\,\text{K}$.}
    \label{fig:v_mag_distribution}
\end{figure}

\begin{table}[tb]
    \setlength{\tabcolsep}{10pt}
    \renewcommand{\arraystretch}{1.5}
    \centering
    \caption{VLEO conditions: orbit velocity $u_{\infty}$, gas temperature $T_{\infty}$, and number density of atomic oxygen $n_{\text{O},\infty}$.}
    \begin{tabular}{c c c}
    \hline\hline
    $u_{\infty} \, / \, (\text{m}/\text{s}) $ & $T_{\infty} \, / \, \text{K} $ & $n_{\text{O},\infty} \, / \, \text{m}^{-3} $ \\
    \hline
     7800 & 934 & $4.698 \cdot 10^{14}$ \\
    \hline\hline
    \end{tabular}
    \label{tab:env_condition}
\end{table}

\subsubsection{Step 1: Data Generation with MD} \label{subsubsec:data_generation}

As already mentioned, MD simulations are computationally expensive, which limits the number of data points that can be generated.
Therefore, it is necessary to specifically choose different incident velocities for the MD simulations.
For efficient selection we, us an approach proposed by the authors in \cite{schuette_2025}, which is introduced in more detail in the following.

As a simplification, we assume that the scattering is not influenced by the incident azimuthal angle $\psi_{\text{i}}$, which is valid for isotropic surfaces, such as amorphous materials, where gas-surface interactions exhibit uniform properties in all directions.
However, this assumption may not apply to anisotropic surfaces, including crystalline materials or engineered surfaces with grooves or structured coatings, where directional dependencies influence scattering behavior.
As a result of this assumption, the scattering is analyzed in the wall frame of reference, as shown in Figure \ref{fig:cos}.
This reference frame is rotated around the surface normal $n$ such that the incident velocity component in the $t2$-direction is zero.
Accordingly, it is only necessary to select different incident velocity magnitudes $v_{\text{i}}$ and polar angles $\theta_{\text{i}}$.

Under VLEO relevant conditions, which are defined in Table \ref{tab:env_condition}, the incident velocity magnitudes are distributed as shown in Figure \ref{fig:v_mag_distribution}.
As can be seen, the distribution can be approximated with a normal distribution.
To select representative incident velocity magnitudes that effectively capture the key characteristics of this distribution, we employ the Gauss-Hermite quadrature.
Gauss-Hermite quadrature approximates integrals of the form:
\begin{equation}
    \int_{-\infty}^{\infty} f(x) e^{-x^2} \, \mathrm{d}x \approx \sum_{s=1}^{n} w_s f(x_s)
\end{equation}
by determining $n$ quadrature points $x_s$ as the roots of the Hermite polynomial $H_n(x)$.
The corresponding weights $w_i$ are derived based on these roots and the properties of the Hermite polynomial \cite{davis_methods_1975}.
Even with a limited number of quadrature points, this method provides an accurate approximation of a normal distribution, ensuring efficient and precise representation of the velocity distribution.
For this study, we set $n=5$ quadrature points.
This choice includes the center point, while also accounting for the tail region.
Using only a small number of points minimizes the required MD simulations, making the process more computationally efficient while still maintaining a high level of approximation accuracy.
The resulting selected incident velocity magnitudes are $6527.8\,\text{m}/\text{s}$, $7261.3\,\text{m}/\text{s}$, $7923.6\,\text{m}/\text{s}$, $8585.9\,\text{m}/\text{s}$, and $9319.4\,\text{m}/\text{s}$.
For the incident polar angles $\theta_{\text{i}}$, we select nine uniformly spaced values between $0.0\,^{\circ}$ and $80.0\,^{\circ}$.
This yields a total of 45 distinct incident velocities, which are then impacted $5000$ times on a surface using MD simulations.
The velocity magnitude of $7923.6\,\text{m}/\text{s}$ across all polar angles is used as validation data for the data-based model, while the remaining magnitudes at all polar angles serve as training data.
Additionally, scattering simulations were performed for velocity magnitudes of $5000.0\,\text{m/s}$ and $11000.0\,\text{m/s}$ at a polar angle of $80\,^{\circ}$ to assess the model's extrapolation performance.

\begin{figure}[tb]
    \centering
    \includegraphics[width=0.8\columnwidth]{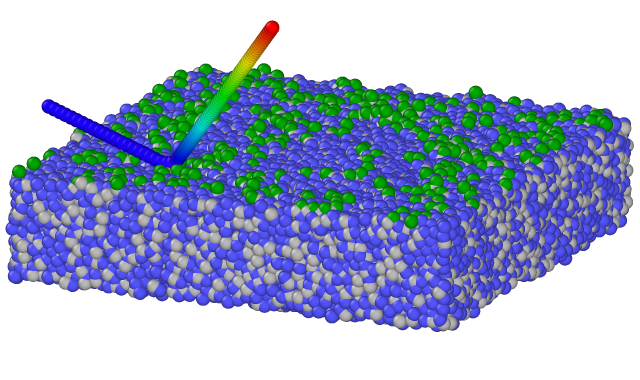}
    \caption{Impingement of atomic oxygen on an amorphous aluminum oxide block ($\text{Al}$: grey, $\text{O}$: blue, adsorbed $\text{O}$: green). The coloring of the reflected atom corresponds to its deviation from the initial plane.}
    \label{fig:md_system}
\end{figure}

For MD simulations, we utilize the IMD package \cite{roth_imp_2019}.
Due to the exposure of atomic oxygen in VLEO, common materials such as aluminum naturally form an oxide layer.
Therefore, the surface material is chosen to be aluminum oxide ($\text{Al}_2\text{O}_3$).
In order to include the effects of atomic roughness, the structure of the surface is modeled as an amorphous block. 
It was generated by melting a surface free structure of $\alpha-\text{Al}_2\text{O}_3$ during a MD simulation with increasing temperature and zero pressure applying periodic boundary conditions.
Subsequently, the melt was cooled by decreasing the temperature, resulting in a solidified amorphous structure containing around $40\,000$ atoms with a surface area of $137\,\text{nm}^2$. 
After creating surfaces by relinquishing the periodic boundary conditions in one direction, additional $550$ oxygen atoms $\text{O}$ were placed on the surface.
This is done to take into account that materials exposed to VLEO adsorb some of the atomic oxygen of the residual atmosphere within short times.
As the precise coverage currently remains uncertain, the selection of $550$ additional $\text{O}$ atoms is somewhat arbitrary. 
Nevertheless, this aspect will be further examined in future investigations.
Figure \ref{fig:md_system} shows the structure of the amorphous aluminum oxide block and the impact of an atomic oxygen particle.
Before starting the impinging simulations, this surface structure was equilibrated to a temperature of $300\,\text{K}$.
This temperature choice is largely due to the absence of direct measurements or comprehensive modeling for this parameter. 
Fortunately, the system's sensitivity to wall temperature variations is generally low \cite{doornbos_2011}, and $300\,\text{K}$ is a commonly used value in the literature.
The interaction between atoms is described using classical analytical potentials based on the Tangney-Scandolo model \cite{tangney_potentials_2002}. 
In this model, oxygen atoms are polarizable, and their dipole moments are calculated self-consistently from the local electric field originating from surrounding charges and dipoles. 
The short range interaction is modeled by a Morse-Stretch potential:

\begin{align} \label{eq:tangney-scandolo_potential}
  U^{\text{MS}}_{ij} &=
	D_{ij}\left\{\exp\left[\gamma_{ij}\left(1-\frac{r_{ij}}{r^0_{ij}}\right)\right] \nonumber \right. \\
    & \left. -2\exp\left[\frac{\gamma_{ij}}{2}\left(1-\frac{r_{ij}}{r^0_{ij}}\right)\right] \right\} 
\end{align}
The model parameters used in this study were calibrated for both bulk and surface structures of $\text{Al}_2\text{O}_3$ \cite{hocker_simulation_2012} and the scattering of neutral atomic oxygen on the surface is investigated.
A simulation time step of $1.018\,\text{fs}$ was employed.
Periodic boundary conditions were applied in the tangential directions.
The incident atomic oxygen particle was initialized at a distance of $>0.1\,\text{nm}$ from the surface to prevent premature interaction before the simulation.
Furthermore, the initial position of the incident particle was randomized.
Impact simulations were performed for a duration of $2\,000$ steps.
The generation of the training and validation data was conducted on 128 cores at the High-Performance Computing Center Stuttgart, requiring nearly four weeks to complete.

\subsubsection{Step 2: Create the conditional Variational Autoencoder} \label{subsubsec:train_cVAE}

\begin{figure}[tb]
    \centering
  \tikzsetnextfilename{cvae_model}%
  \input{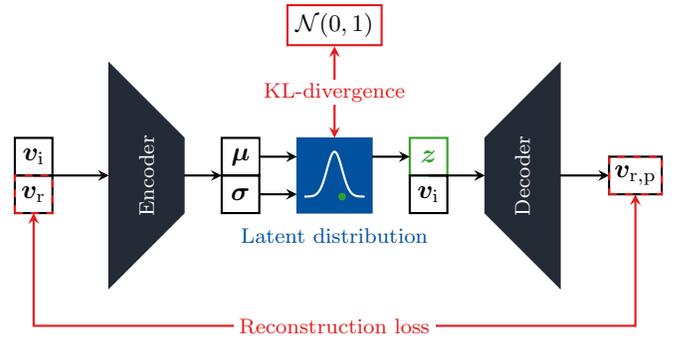}%

    \caption{Architecture of the conditional Variational Autoencoder (cVAE).}
    \label{fig:cvae_model}
\end{figure}

The cVAE belongs to the family of deep generative models.
The traditional Variational Autoencoder (VAE) framework was introduced by Diederik Kingma and Max Welling in 2013 \cite{kingma2022autoencodingvariationalbayes}.
This framework was enhanced by incorporating a condition into the encoder and decoder inputs, which was popularized by Sohn et al. in 2015 \cite{Sohn_cvae_2015}.
As a result, the model generates outputs that are tailored to a specific condition.
The architecture of the cVAE comprises an encoder followed by a decoder.
The encoder parameterizes a probability distribution (typically a normal distribution) by outputting the mean and variance of the distribution.
This procedure effectively compresses the input into a conditional distribution in the latent space.
A latent vector $\bm{z}$ is then sampled from this distribution and fed into the decoder along with the condition.
With the latent sample and the condition as input, the decoder generates the output.
The overall loss function of a VAE consists of two parts.
First, the reconstruction loss, which is commonly the mean squared error between the input and the predicted output of the decoder.
It quantifies the fidelity of the decoded output relative to the original input.
Second, the Kullback-Leibler divergence between the conditional distribution defined by the encoder and the predefined prior distribution, usually the standard normal distribution.
Minimizing this divergence ensures that the latent representations remain well-structured and that the aggregate distribution over $\bm{z}$ conforms to the prior.

In our case, the condition is the incident velocity $\bm{v}_{\text{i}}$ and the output we want to generate is the reflected velocity $\bm{v}_{\text{r}}$.
Once trained, the derived decoder of the cVAE can be used as the scattering kernel $\mathcal{K}(\bm{v}_{\text{i}} \rightarrow \bm{v}_{\text{r}})$ in the DSMC simulation, enabling the determination of the reflected velocity $\bm{v}_{\text{r}}$ based on the incident velocity $\bm{v}_{\text{i}}$.
Both the encoder and decoder are typically implemented as fully connected feedforward neural networks.
The principle of the cVAE is illustrated in Figure \ref{fig:cvae_model}.

For constructing the cVAE, we utilize the Keras library \cite{chollet_keras_2015} with TensorFlow \cite{tensorflow2015-whitepaper} as backend.
Various architectures involving different numbers of hidden layers, neurons, activation functions, and latent space dimensions were tested.
The configuration detailed below exhibited the best performance in terms of training and validation metrics.

\begin{figure}[tb]
    \centering
  \tikzsetnextfilename{softplus}%
  \input{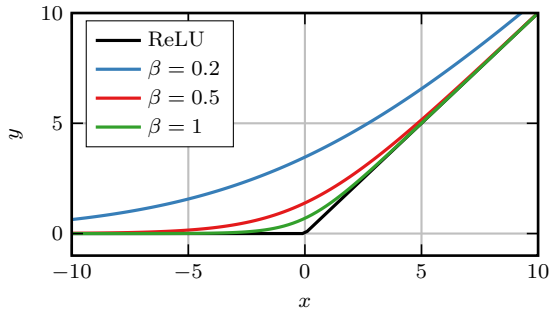}%

    \caption{Comparison of the Softplus activation function with different $\beta$ and the ReLU activation function.}
    \label{fig:softplus}
\end{figure}

The encoder has an input layer with six neurons, which represent the incident velocity $\bm{v}_{\text{i}}$ and one of the respective reflected velocities $\bm{v}_{\text{r}}$.
Two hidden layers with $64$ and $32$ neurons are selected.
Both hidden layers use the Exponential Linear Unit (ELU) activation function \cite{clevert_elu_2016}, which performed better compared to other activation functions, like the Rectified Linear Unit (ReLU) \cite{nair_relu_2010} or leaky ReLU \cite{maas_leaky_2013}, in our tests.
The dimension of the latent space is set to three, and a normal distribution with a diagonal covariance matrix is used.
Sampling from this distribution is much easier, simplifying backpropagation.
Moreover, KL-divergence then offers a closed-form solution \cite{kingma2022autoencodingvariationalbayes}.
Consequently, the encoder has an output layer with six neurons, three for the mean and three for the standard deviation of the latent space.
From this distribution a sample $\bm{z}$ is drawn.
The decoder takes the sample $\bm{z}$ and the incident velocity $\bm{v}_{\text{i}}$ as input, concatenates them, and passes them through two hidden layers with $32$ and $64$ neurons with ELU activation functions.
The output layer of the decoder is twofold:
The first part predicts the tangential reflected velocity along the $t1$-direction and the $t2$-direction using a linear activation function, and the second part predicts the normal reflected velocity through a Softplus activation function \cite{dugas_softlus_2000}.
The Softplus activation function is defined as
\begin{equation}
    y(x;\beta) = \frac{1}{\beta} \log \left(1 + \exp[\beta x] \right)
\end{equation}
and it is shown in Figure \ref{fig:softplus} for different values of $\beta$ in comparison to the ReLU activation function.
The Softplus activation function guarantees positive normal reflected velocities.
Unlike the ReLU activation function, Softplus prevents saturation for negative values. 
ReLU's saturation issue can result in an excessive number of particles with a normal velocity of zero, especially at incident velocity magnitudes lower than those in the training data. 
As a result, Softplus provides better performance when extrapolating to lower velocity magnitudes.
Among the different configurations, Softplus with $\beta=0.2$ demonstrated the best performance, providing stable and accurate results.

The loss function is formulated as the sum of two terms: (i) the mean squared error between the predicted and the true reflected velocity, and (ii) the Kullback-Leibler divergence between the prior, the standard normal distribution, and the posterior distribution.
Model training is performed using the Adam optimizer in $100$ epochs with a batch size of $32$.
An initial learning rate of $10^{-3}$ is employed.
Then a step decay schedule is implemented, in which the learning rate is reduced by a factor of $10^{-1}$ every $20$ epoch.
A higher initial learning rate helps the model to quickly explore a wide range of parameter values and identify promising regions in the parameter space.
This rapid exploration prevents the optimization process from getting stuck in a local minima.
Once the model has roughly located a good region, progressively decreasing the learning rate allows for finer adjustments.
This gradual refinement minimizes the risk of overshooting the optimal point and promotes a steadier, more reliable convergence towards the best solution.
The result of the training process as well as the performance evaluation of the cVAE on the validation and extrapolation data is discussed in Section \ref{subsec:cvae_evaluation}.

\section{Data and Model Evaluation} \label{sec:data_and_model_evaluation}

In the following Sections, we first discuss the derived training data from the MD scattering simulations.
Afterwards, the performance of the cVAE is evaluated on the training, validation, and extrapolation data.

\subsection{Analysis of the Training Data} \label{subsec:md_results}

The MD simulations determines the trajectories of the impinging particles.
There are two possible mechanisms which can occur after the interaction with the surface, namely adsorption or reflection.
In order to determine the occurring process for each impingement simulation, we use the following energy-based criterion. A particle is identified as reflected if
\begin{align}
	|E_{\text{pot}}| < 0.001\,\mathrm{eV} \;\;&\wedge\;\; E_{\text{pot}}+\frac{1}{2}mv_{\text{n}}^2 > 0\,\mathrm{eV}\nonumber \\ \;\;&\wedge\;\; v_{\text{n}} > 0\, \text{m/s}
\end{align}
whereas it is identified as adsorbed if
\begin{equation}
	E_{\text{pot}} \le -0.001\,\mathrm{eV} \quad\wedge\quad E_{\text{pot}}+\frac{1}{2}mv_{\text{n}}^2 \le 0\,\mathrm{eV}.
\end{equation}
In some rare cases, the impinging particle meets neither of these criteria after the $2000$ simulation steps performed.
These cases are classified as ``unclear''.
The reflection criterion has the advantage that particles classified as ``reflected'' are outside the force fields of the other atoms, resulting in straight trajectories and therefore meaningful reflection angles.

\begin{figure*}[tb]
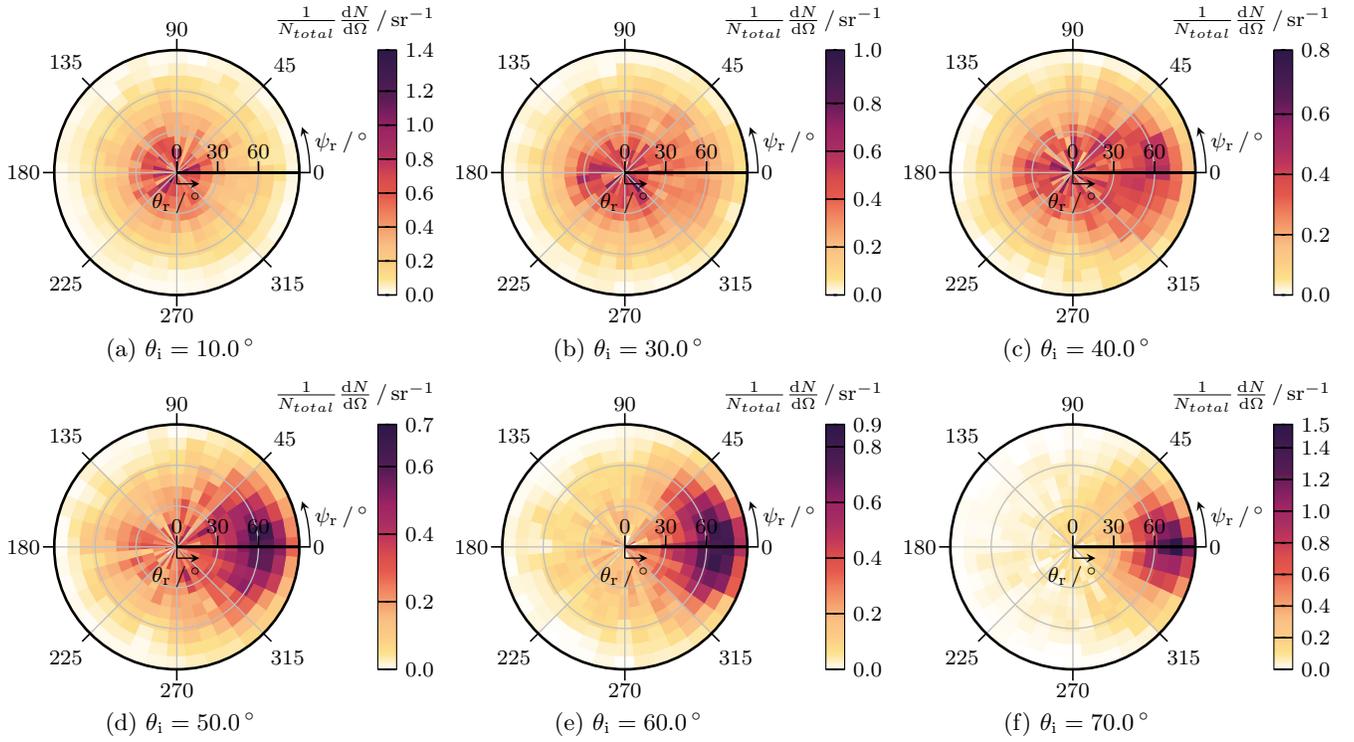

    \def\subfigwidth{0.27\textwidth}
    \centering
    \begin{subfigure}[t]{\subfigwidth}
        \centering
  \tikzsetnextfilename{nbr_angular_dist_v_mag_7923.6_10.0}%
  \input{sections/results/data_and_model_evaluation/md_results/tikzpictures/nbr_angular_dist/nbr_angular_dist_v_mag_7923.6_10.0.tex}%

        \vspace{-5mm}
        \caption{$\theta_{\text{i}} = 10.0\,^{\circ}$}
        \label{subfig:nbr_angular_dist_10_deg}
    \end{subfigure}
    \hspace{9mm}
    \begin{subfigure}[t]{\subfigwidth}
        \centering
  \tikzsetnextfilename{nbr_angular_dist_v_mag_7923.6_30.0}%
  \input{sections/results/data_and_model_evaluation/md_results/tikzpictures/nbr_angular_dist/nbr_angular_dist_v_mag_7923.6_30.0.tex}%

        \vspace{-5mm}
        \caption{$\theta_{\text{i}} = 30.0\,^{\circ}$}
        \label{subfig:nbr_angular_dist_30_deg}
    \end{subfigure}
    \hspace{9mm}
    \begin{subfigure}[t]{\subfigwidth}
        \centering
  \tikzsetnextfilename{nbr_angular_dist_v_mag_7923.6_40.0}%
  \input{sections/results/data_and_model_evaluation/md_results/tikzpictures/nbr_angular_dist/nbr_angular_dist_v_mag_7923.6_40.0.tex}%

        \vspace{-5mm}
        \caption{$\theta_{\text{i}} = 40.0\,^{\circ}$}
        \label{subfig:nbr_angular_dist_40_deg}
    \end{subfigure}
    \hspace{9mm}

    \begin{subfigure}[t]{\subfigwidth}
        \centering
  \tikzsetnextfilename{nbr_angular_dist_v_mag_7923.6_50.0}%
  \input{sections/results/data_and_model_evaluation/md_results/tikzpictures/nbr_angular_dist/nbr_angular_dist_v_mag_7923.6_50.0.tex}%

        \vspace{-5mm}
        \caption{$\theta_{\text{i}} = 50.0\,^{\circ}$}
        \label{subfig:nbr_angular_dist_50_deg}
    \end{subfigure}
    \hspace{9mm}
    \begin{subfigure}[t]{\subfigwidth}
        \centering
  \tikzsetnextfilename{nbr_angular_dist_v_mag_7923.6_60.0}%
  \input{sections/results/data_and_model_evaluation/md_results/tikzpictures/nbr_angular_dist/nbr_angular_dist_v_mag_7923.6_60.0.tex}%

        \vspace{-5mm}
        \caption{$\theta_{\text{i}} = 60.0\,^{\circ}$}
        \label{subfig:nbr_angular_dist_60_deg}
    \end{subfigure}
    \hspace{9mm}
    \begin{subfigure}[t]{\subfigwidth}
        \centering
  \tikzsetnextfilename{nbr_angular_dist_v_mag_7923.6_70.0}%
  \input{sections/results/data_and_model_evaluation/md_results/tikzpictures/nbr_angular_dist/nbr_angular_dist_v_mag_7923.6_70.0.tex}%

        \vspace{-5mm}
        \caption{$\theta_{\text{i}} = 70.0\,^{\circ}$}
        \label{subfig:nbr_angular_dist_70_deg}
    \end{subfigure}
    \hspace{9mm}

    \caption{Angular distribution of the reflected particle count $\mathrm{d}N$ with regard to the total reflected particles $N_{\text{t}}$ and the solid angle $\mathrm{d}\Omega$ for an incident particle velocity magnitude of $|\bm{v}_{\text{i}}| = 7923.6 \, \mathrm{m/s}$.}
    \label{fig:nbr_angular_distribution_md}
\end{figure*}

\begin{figure}[tb]
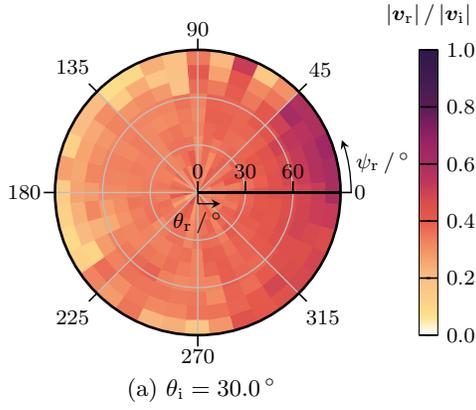

    \centering
    \begin{subfigure}[t]{0.3\textwidth}
        \centering
  \tikzsetnextfilename{rel_vel_dist_v_mag_7923.6_30.0}%
  \input{sections/results/data_and_model_evaluation/md_results/tikzpictures/velo_dist/rel_vel_dist_v_mag_7923.6_30.0.tex}%

        \vspace{-5mm}
        \caption{$\theta_{\text{i}} = 30.0\,^{\circ}$}
        \label{subfig:velo_dist_7923.6_30_deg}
    \end{subfigure}
    \quad
    \begin{subfigure}[t]{0.3\textwidth}
        \centering
  \tikzsetnextfilename{rel_vel_dist_v_mag_7923.6_70.0}%
  \input{sections/results/data_and_model_evaluation/md_results/tikzpictures/velo_dist/rel_vel_dist_v_mag_7923.6_70.0.tex}%

        \vspace{-5mm}
        \caption{$\theta_{\text{i}} = 70.0\,^{\circ}$}
        \label{subfig:velo_dist_7923.6_70_deg}
    \end{subfigure}

    \caption{Angular distribution of the reflected velocity magnitude $|\bm{v}_{\text{r}}|$ of the particle with regard to the incident particle velocity magnitude $|\bm{v}_{\text{i}}| = 7923.6\,\text{m/s} $ for different incident polar angels $\theta_{\text{i}}$.}
    \label{fig:velo_dist_7923.6_angle_comp_md}
\end{figure}

\begin{figure}[tb]
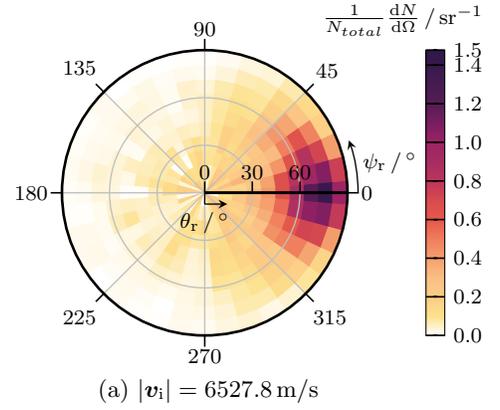

    \centering
    \begin{subfigure}[t]{0.3\textwidth}
        \centering
  \tikzsetnextfilename{nbr_angular_dist_v_mag_6527.8_70.0}%
  \input{sections/results/data_and_model_evaluation/md_results/tikzpictures/nbr_angular_dist/nbr_angular_dist_v_mag_6527.8_70.0.tex}%

        \vspace{-5mm}
        \caption{$|\bm{v}_{\text{i}}| = 6527.8\,\text{m/s}$}
        \label{subfig:nbr_angular_dist_6527.8_70_deg}
    \end{subfigure}
    \quad
    \begin{subfigure}[t]{0.3\textwidth}
        \centering
  \tikzsetnextfilename{nbr_angular_dist_v_mag_9319.4_70.0}%
  \input{sections/results/data_and_model_evaluation/md_results/tikzpictures/nbr_angular_dist/nbr_angular_dist_v_mag_9319.4_70.0.tex}%

        \vspace{-5mm}
        \caption{$|\bm{v}_{\text{i}}| = 9319.4\,\text{m/s}$}
        \label{subfig:nbr_angular_dist_9319.4_70_deg}
    \end{subfigure}

    \caption{Angular distribution of the reflected count $\mathrm{d}N$ with regard to the total reflected particles $N_{\text{t}}$ and the solid angle $\mathrm{d}\Omega$ for an incident polar angel of $\theta_{\text{i}} = 70\,^{\circ}$ and different velocity magnitudes $|\bm{v}_{\text{i}}|$.}
    \label{fig:nbr_angular_dist_70_deg_velo_mag_comp_md}
\end{figure}

Since the angular distribution of the particle counts and the angular distribution of the velocity for the reflected particles are very important for drag estimation, we will show and discuss the MD results of these distributions in this Section.
For this task, we subdivide the parameter space into $10\,^{\circ}$ intervals (``bins'').
The angular distributions of the particle counts are given as the fraction of particles counted in the bins $\mathrm{d}N$ and the product of the total count of reflected particles $N_{\text{t}}$ and the solid angle corresponding to the bin $\mathrm{d}\Omega = \Delta\psi_{\text{r}}  \left(\cos(\theta_{\text{r}}) - \cos(\theta_{\text{r}}+\Delta\theta_{\text{r}})\right)$.
This choice leads to a constant value for a random distribution of reflection angles and to high particle counts for low polar angles in the case of diffuse scattering.
The velocities of the reflected particles are evaluated as the fraction of the average velocity magnitude in the bin and the initial velocity magnitude of the impinging atom.
The simulation results for the particle count and velocity distributions are visualized by polar plots depending on the polar angle $\theta_{\text{r}}$ and the azimuthal angle $\psi_{\text{r}}$ in Figures \ref{fig:nbr_angular_distribution_md}, \ref{fig:velo_dist_7923.6_angle_comp_md}, \ref{fig:nbr_angular_dist_70_deg_velo_mag_comp_md}.

As shown in Figure \ref{fig:nbr_angular_distribution_md}, for the initial velocity magnitude of $7923.6\,\text{m/s}$, the angular distribution of the reflected particle count strongly depends on the angle of incidence.
A high ratio of specular reflection is observed for high angles of incidence.
For $\theta_{\text{i}}=70\,^{\circ}$ (Figure \ref{subfig:nbr_angular_dist_70_deg}) there is a pronounced maximum between $60\,^{\circ}<\theta_{\text{r}} < 80\,^{\circ}$ and $\psi_{\text{r}}=0\,^{\circ}$.
A clear maximum in the angular distributions of the reflected particle count at the corresponding incidence angle can also be seen in the distributions for $\theta_{\text{i}}=50\,^{\circ}$ and $\theta_{\text{i}}=60\,^{\circ}$ (Figures \ref{subfig:nbr_angular_dist_50_deg} and \ref{subfig:nbr_angular_dist_60_deg}).
However, the angular distributions are wider with respect to both $\theta_{\text{r}}$ and $\psi_{\text{r}}$ and the amount of diffuse scattering increases compared to higher incident polar angles $\theta_{\text{i}}$.
A further decrease of the angle of incidence leads to an even higher ratio of diffuse reflection.
At $\theta_{\text{i}}=10\,^{\circ}$ and $\theta_{\text{i}}=30\,^{\circ}$ (Figures \ref{subfig:nbr_angular_dist_10_deg} and \ref{subfig:nbr_angular_dist_30_deg}) we observe pronounced maxima for low $\theta_{\text{r}}$ as expected for diffuse scattering.

Reflection and adsorption rates were found to be correlated with the angular
distributions of reflected particles.
As demonstrated in Table \ref{tab:refl}, the reflection rates increase and the adsorption rates decrease with increasing angle of incidence.

The reflected angular velocity distribution of the reflected particles shows a clear dependence of the velocities on the reflection angles $\theta_{\text{r}}$ and $\psi_{\text{r}}$.
As seen in Figure \ref{fig:velo_dist_7923.6_angle_comp_md}, high particle velocities are obtained for high positive values of $\theta_{\text{r}}$ in a wide range of $\psi_{\text{r}}$.
This trend is observed for all angles of incidence even in the cases where mostly diffuse scattering is observed, e.g. for $\theta_{\text{i}}=30\,^{\circ}$ in Figure \ref{subfig:velo_dist_7923.6_30_deg}. However,
the absolute value of the velocity of the fastest particles is different. For
high angles of incidence, they retain their initial velocity
(Figure \ref{subfig:velo_dist_7923.6_70_deg}) whereas the final velocities decrease for lower angles of incidence, e.g. to $0.7\,\cdot\, |\bm{v}_{\text{i}}|$ for $\theta_{\text{i}}=30\,^{\circ}$
(Figure \ref{subfig:velo_dist_7923.6_30_deg}).

\begin{table*}[tb]
\centering
\renewcommand{\arraystretch}{1.2}
\caption{Rates of the observed oxygen-surface interaction processes for
  different angles of incidence and initial velocities.}
\label{tab:refl}
\begin{tabular}{p{3cm}p{1.1cm}p{1.1cm}p{1.1cm}p{1.1cm}p{1.1cm}p{1.1cm}p{1.1cm}p{1.1cm}p{1.1cm}}
    \hline\hline
    $|\bm{v}_{\text{i}}|=6527.8\,\text{m/s}$ & $0\,^{\circ}$ & $10\,^{\circ}$ & $20\,^{\circ}$ & $30\,^{\circ}$ & $40\,^{\circ}$  & $50\,^{\circ}$ & $60\,^{\circ}$ & $70\,^{\circ}$ & $80\,^{\circ}$\\
\hline
reflected $\,/\,\%$ & 87.42 & 88.92 & 89.02 & 89.6 & 90.72 & 92.76 & 93.4 & 95.78 & 94.02 \\
adsorbed $\,/\,\%$ & 10.6 & 9.5 & 9.62 & 8.68 & 7.9 & 5.9 & 5.06 & 2.96 & 3.42 \\
unclear $\,/\,\%$ & 1.98 & 1.58 & 1.36 & 1.72 & 1.38 & 1.34 & 1.54 & 1.26 & 2.56 \\
  \hline
  $|\bm{v}_{\text{i}}|=7923.6\,\text{m/s}$ & $0\,^{\circ}$ & $10\,^{\circ}$ & $20\,^{\circ}$ & $30\,^{\circ}$ & $40\,^{\circ}$  & $50\,^{\circ}$ & $60\,^{\circ}$ & $70\,^{\circ}$ & $80\,^{\circ}$\\
\hline
reflected $\,/\,\%$ & 89.36 & 90.4 & 90.54 & 91.1 & 91.66 & 94.54 & 96.06 & 96.54 & 96.46 \\
adsorbed $\,/\,\%$ & 9.42 & 8.3 & 8.38 & 7.58 & 7.26 & 4.38 & 3.04 & 2.56 & 1.84 \\
unclear $\,/\,\%$ & 1.22 & 1.3 & 1.08 & 1.32 & 1.08 & 1.08 & 0.9 & 0.9 & 1.7 \\
  \hline
  $|\bm{v}_{\text{i}}|=9319.4\,\text{m/s}$ & $0\,^{\circ}$ & $10\,^{\circ}$ & $20\,^{\circ}$ & $30\,^{\circ}$ & $40\,^{\circ}$  & $50\,^{\circ}$ & $60\,^{\circ}$ & $70\,^{\circ}$ & $80\,^{\circ}$\\
\hline
reflected $\,/\,\%$ & 89.92 & 90.86 & 91.18 & 91.88 & 93.44 & 94.82 & 96.04 & 97.4 & 97.26 \\
adsorbed $\,/\,\%$ & 8.86 & 8.2 & 7.78 & 7.32 & 5.8 & 4.46 & 3.34 & 2.02 & 1.82 \\
unclear $\,/\,\%$ & 1.22 & 0.94 & 1.04 & 0.8 & 0.76 & 0.72 & 0.62 & 0.58 & 0.92 \\
  \hline  \hline
\end{tabular}
\end{table*}

In contrast to the strong dependency of the angular particle count distribution and the angular velocity distribution on the incident polar angle, the initial velocity has a minor influence.
We observe the same trends for all initial velocities in the considered range from $6527.8\,\text{m/s}$ to $9319.4\,\text{m/s}$.
The only difference is that the angular distributions are slightly broader for lower velocities (Figure \ref{fig:nbr_angular_dist_70_deg_velo_mag_comp_md}) and the reflection rates are slightly decreased (Table \ref{tab:refl}).

Unfortunately, experimental results that allow to validate the simulation results are rare.
There are measurements \cite{minton} of atomic oxygen reflection on sapphire and amorphous alumina layers on polymers with an angle
of incidence $\theta_{\text{i}}=70\,^{\circ}$ and $|\bm{v}_{\text{i}}| \approx 7900\,\text{m/s}$.
In accordance with our simulations they  obtain a pronounced maximum at the reflection angle $\theta_{\text{r}}=70^{\circ}$.
Furthermore, the experiments also revealed that the atoms reflected at high angles ($\theta_{\text{r}}>70\,^{\circ}$) retain their kinetic energy.
The average kinetic energy of the particles reflected at $\theta_{\text{r}}=40\,^{\circ}$ is about half of the initial energy.
This corresponds to a velocity of $0.7\,\cdot\, |\bm{v}_{\text{i}}|$ which agrees well with our simulations (Figure \ref{subfig:velo_dist_7923.6_70_deg}).

\subsection{Validation of the cVAE Model} \label{subsec:cvae_evaluation}

\begin{figure}[tb]
    \centering
  \tikzsetnextfilename{losses}%
  \input{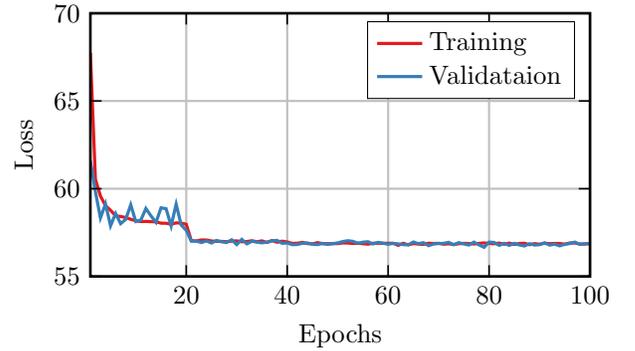}%

    \caption{Training and validation losses of the cVAE.}
    \label{fig:losses}
\end{figure}

\begin{figure*}[tb]
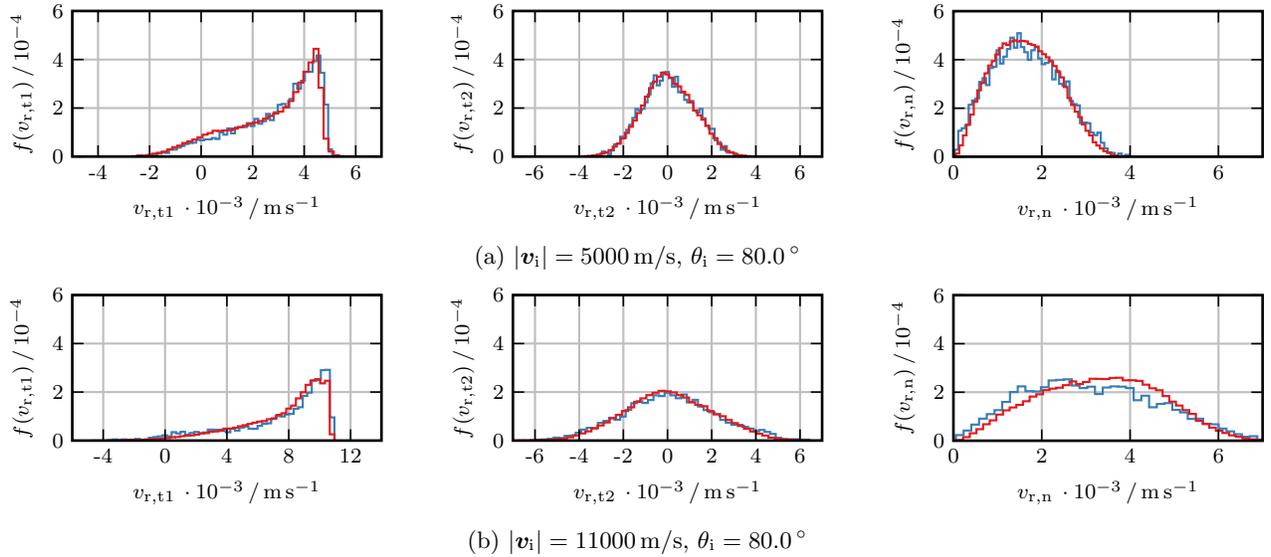

    \centering
    \begin{minipage}{0.99\textwidth}
        \centering
        \begin{minipage}{0.324\textwidth}
            \centering
  \tikzsetnextfilename{v_mag_5000.0_theta_80.0_t1_directions}%
  \input{sections/appendix/tikzpictures/histogram_v_mag_5000.0_theta_80.0/v_mag_5000.0_theta_80.0_t1_directions.tex}%

        \end{minipage}
        \begin{minipage}{0.324\textwidth}
            \centering
  \tikzsetnextfilename{v_mag_5000.0_theta_80.0_t2_directions}%
  \input{sections/appendix/tikzpictures/histogram_v_mag_5000.0_theta_80.0/v_mag_5000.0_theta_80.0_t2_directions.tex}%

        \end{minipage}
        \begin{minipage}{0.324\textwidth}
            \centering
  \tikzsetnextfilename{v_mag_5000.0_theta_80.0_n_directions}%
  \input{sections/appendix/tikzpictures/histogram_v_mag_5000.0_theta_80.0/v_mag_5000.0_theta_80.0_n_directions.tex}%

        \end{minipage}
        \subcaption{$|\bm{v}_{\text{i}}| = 5000 \, \text{m}/\text{s}$, $\theta_{\text{i}} = 80.0 \, ^\circ$}
        \label{fig:histogram_extrapolation_lower}
    \end{minipage}
    \begin{minipage}{0.99\textwidth}
        \centering
        \begin{minipage}{0.324\textwidth}
            \centering
  \tikzsetnextfilename{v_mag_11000.0_theta_80.0_t1_directions}%
  \input{sections/appendix/tikzpictures/histogram_v_mag_11000.0_theta_80.0/v_mag_11000.0_theta_80.0_t1_directions.tex}%

        \end{minipage}
        \begin{minipage}{0.324\textwidth}
            \centering
  \tikzsetnextfilename{v_mag_11000.0_theta_80.0_t2_directions}%
  \input{sections/appendix/tikzpictures/histogram_v_mag_11000.0_theta_80.0/v_mag_11000.0_theta_80.0_t2_directions.tex}%

        \end{minipage}
        \begin{minipage}{0.324\textwidth}
            \centering
  \tikzsetnextfilename{v_mag_11000.0_theta_80.0_n_directions}%
  \input{sections/appendix/tikzpictures/histogram_v_mag_11000.0_theta_80.0/v_mag_11000.0_theta_80.0_n_directions.tex}%

        \end{minipage}
        \subcaption{$|\bm{v}_{\text{i}}| = 11000 \, \text{m}/\text{s}$, $\theta_{\text{i}} = 80.0 \, ^\circ$}
        \label{fig:histogram_extrapolation_higher}
    \end{minipage}
    \caption{Probability density functions of the conditional Variational Autoencoder (\textcolor{customred}{\rule[0.5ex]{0.5cm}{1pt}}) and the Molecular Dynamics simulation (\textcolor{customblue}{\rule[0.5ex]{0.5cm}{1pt}}) for extrapolation to an incident velocity magnitude of $5000\,\text{m}/\text{s}$ and $11000\,\text{m}/\text{s}$ at an incident angle of $80.0\,^{\circ}$.}
    \label{fig:histograms_extrapolation}
\end{figure*}

The cVAE is trained using the reflected velocity distributions previously obtained from MD simulations for the corresponding incident velocity vectors.
However, the velocity distributions for the incident velocity magnitude $7923.6\,\text{m}/\text{s}$ with the incident angles $\theta_{\text{i}}$ from $0.0\,^{\circ}$ to $80.0\,^{\circ}$ are used for validation.
The model training was performed on an AMD Ryzen 5 Pro 7530u CPU and took roughly ten minutes, but it could be significantly reduced with the use of a GPU.
The loss on the training and validation data was monitored during training and is shown in Figure \ref{fig:losses}.
The rapid decrease of the training and validation loss at $20$ epochs is due to the reduced learning rate from $10^{-3}$ to $10^{-4}$ at this point.
This indicates that the model is able to get closer to the minimum of the loss function, while a higher learning rate would have led to oscillations around the minimum.
Additionally, both the validation and training losses decrease over epochs, indicating that the model is effectively learning generalizable patterns rather than overfitting to the training data. 
In contrast, overfitting would be characterized by a continued decrease in training loss while the validation loss begins to rise, suggesting that the model is memorizing training data rather than adapting to unseen inputs.
The probability density functions of the cVAE and the MD simulation for the validation data are shown in Figures \ref{fig:histograms_md_cVAE_0_40} to \ref{fig:histograms_md_cVAE_50_80} in the appendix.
For the histograms of the cVAE $100\,000$ samples were generated, while for the MD simulation, the number of samples are between $4000$ and $5000$, as not all impacts resulted in a reflection.
This was previously demonstrated in Table \ref{tab:refl}.
The cVAE demonstrates a remarkable ability to replicate the results of the MD simulation.
In particular, it effectively captures the observed progression in the $t2$-direction. The distribution transitions from a normal profile at lower angles to a right-skewed form at moderate angles, evolves into a distribution characterized by negative kurtosis at higher angles, and finally becomes left-skewed at the highest angles.
Similar behavior is observed across all other incident velocity magnitudes.
Moreover, despite the presence of some noise in the MD simulation data, the cVAE model effectively identifies underlying patterns, ensuring a reliable generalization.
The model is also capable of interpolating between different incident velocities, which can be seen on the validation data in Figures \ref{fig:histograms_md_cVAE_0_40} to \ref{fig:histograms_md_cVAE_50_80}.
To investigate the extrapolation ability to incident velocity magnitudes outside the training range, additional MD simulations were performed for an incident velocity magnitude of $5000\,\text{m}/\text{s}$ and $11000\,\text{m}/\text{s}$ both for an incident polar angle of $80\,^{\circ}$.
The reflected velocity distributions are shown in Figures \ref{fig:histogram_extrapolation_lower} and \ref{fig:histogram_extrapolation_higher}.
Even though the cVAE model experiences a slight decrease in accuracy for the normal reflected velocity distribution at an incident velocity magnitude of $11000\,\text{m}/\text{s}$, it still demonstrates impressive performance. 
This is especially notable given how significantly these velocity magnitudes deviate from the incident training velocity magnitudes.
As a result, the cVAE can still be used for different bulk velocity $u_{\infty}$ and gas temperature $T_{\infty}$ in the VLEO regime.
Nevertheless, extrapolation to incident particle velocities that lie well outside the trained range is discouraged.
Incorporating additional data points for incident velocities could potentially further enhance the model's fidelity and generalization ability.
After training and validation, the decoder is implemented in the DSMC method of the open source simulation program PICLas, which was and is being developed at the Institute of Space Systems \cite{fasoulas_combining_2019}.

In summary, we directly derive the scattering kernels $\mathcal{K}(\bm{v}_{\text{i}} \rightarrow \bm{v}_{\text{r}})$ for the respective $45$ incident velocity vectors $\bm{v}_{\text{i}}$ from the MD simulations.
Afterwards, the cVAE is trained on these scattering kernels, which allows us to generate the scattering kernels for any incident velocity vector $\bm{v}_{\text{i}}$.
In total, $180\,000$ impacts for the generation of the training data were simulated using MD, corresponding to $36$ ensembles (four velocity magnitudes with nine different polar angles) of $5000$ impacts each.
In contrast, alternative data-driven models, such as the GMM or DET, require $100\,000$ impacts for each distinct bulk polar angle.
Therefore, selecting nine different bulk polar angles would result in nearly one million impacts.
Moreover, the resulting GMM or DET is only valid for the specific bulk velocity and gas temperature conditions, whereas the derived cVAE can be used for different conditions in the VLEO regime.

It is important to note, that the training process for the cVAE model relied exclusively on non-equilibrium data characterized by the very high velocities typical of VLEO conditions.
As a result, the model demonstrates robust performance within these non-equilibrium regimes.
However, it is important to note that because the model was tailored to these high velocities and non-equilibrium conditions, its predictive capability may be limited when applied to scenarios characterized by near-equilibrium conditions and low velocities.
In such cases, the model may not accurately reproduce the expected equilibrium behavior because it was not exposed to, nor trained on, data representative of these conditions.
To address this limitation and improve model applicability for near-equilibrium or low-speed problems, a logical step would be to enrich the training dataset with MD simulation data that spans a wider range of conditions, including those representative of near-equilibrium and low incident velocities.
Additionally, a term can be incorporated into the loss function to penalize any deviation from the expected variance of the reflected thermal velocities, aligning it with the thermal velocity determined by the wall temperature.
Future work will focus on validating these modifications and extending the model's domain.

\section{Application of the cVAE model} \label{sec:application_cvae}

In the following two Sections, we compare the previously derived cVAE scattering model with the most common DSMC scattering kernel, the Maxwell model.
The comparison is done on the basis of the aerodynamic coefficients and the angular distribution of the molecular flux, which are crucial for the design of satellites in VLEO.
In both cases, we consider an amorphous $\text{Al}_2\text{O}_3$ surface with no roughness on the micro- or nanometer scale and a wall temperature of $T_{\text{w}} = 300 \, \text{K}$ under VLEO conditions, as shown in Table \ref{tab:env_condition}.

\subsection{Aerodynamic Coefficients} \label{subsec:aerodynamic_coefficients}

\begin{figure}[tb]
    \centering
  \tikzsetnextfilename{flat_plate_sketch}%
  \input{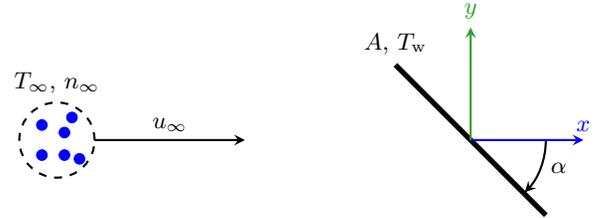}%

    \caption{Schematic illustration of a flat plate with a temperature of $T_{\text{w}}$, an area $A$, and an angle of attack $\alpha$ relative to the incident particles with bulk velocity $u_\infty$, number density $n_{\infty}$ and temperature $T_\infty$.}
    \label{fig:flat_plate_sketch}
\end{figure}

In order to compare the aerodynamic performance of the cVAE model and the Maxwell model, the behavior of a flat plate is studied. Figure \ref{fig:flat_plate_sketch} illustrates the setup of the flat plate with an area $A$ inclined at an angle of attack $\alpha$, with regard to the bulk velocity $u_\infty$ of the incident particles.
The particles are distributed according to the temperature $T_\infty$ and have a number density $n_{\infty}$.

Assuming free molecular flow, the Maxwell model provides an analytical expression for the aerodynamic coefficients of the flat plate.
The lift coefficient $C_{\ell}$ and the drag coefficient $C_{\text{d}}$ derived by Bird \cite{bird_molecular_1994}, are given by equations \eqref{eq:c_l_bird} and \eqref{eq:c_d_bird}, respectively.
\begin{align} \label{eq:c_l_bird}
    C_{\ell} &= \frac{4 \varepsilon}{\sqrt{\pi} s} \sin(\alpha) \cos(\alpha) \exp\left( -s^2 \sin^2(\alpha) \right) \notag \\
    &+ \frac{\cos(\alpha)}{s^2} \left\{ 1+ \varepsilon \left(1 + 4 s^2 \sin^2(\alpha) \right) \right\} \cdot \erf(s \sin(\alpha)) \notag \\
    &+ \frac{1 - \varepsilon}{s} \sqrt{\pi} \sin(\alpha) \cos(\alpha) \sqrt{\frac{T_\text{r}}{T_{\infty}}}
\end{align}

\begin{align} \label{eq:c_d_bird}
    C_{\text{d}} &= 2 \frac{1 - \varepsilon \cos(2\alpha)}{\sqrt{\pi} s} \exp\left( -s^2 \sin^2(\alpha) \right) \notag \\
    &+ \frac{\sin(\alpha)}{s^2} \left[ 1 + 2 s^2 + \varepsilon \left\{1 - 2 s^2 \cos(2\alpha)\right\} \right] \erf(s \sin(\alpha)) \notag \\
    &+ \frac{1 - \varepsilon}{s} \sqrt{\pi} \sin^2(\alpha) \sqrt{\frac{T_{\text{r}}}{T_{\infty}}}
\end{align}
Here, $\varepsilon$ is the fraction of specular reflection.
In comparison, the accommodation coefficient $\sigma$ of the Maxwell scattering kernel \eqref{eq:maxwell_kernel} is defined as the fraction of molecules that are diffusely reflected.
Therefore, the two parameters are related by $\varepsilon = 1 - \sigma$.
The molecular speed ratio, $s$, is given by $s = u_{\infty} / \sqrt{2 k T_{\infty} / m}$, where $u_{\infty}$ is the bulk velocity of the gas, $k$ is the Boltzmann constant, and $m$ is the mass of atomic oxygen ($2.657 \cdot 10^{-26}\,\text{kg}$).
$T_{\text{r}}$ and $T_{\infty}$ are the temperatures of the reflected particles and the gas, respectively.
In this study, the temperature of the reflected particles $T_{\text{r}}$ is set equal to the wall temperature $T_{\text{w}}$.

The lift and drag coefficients of the cVAE scattering model are obtained by DSMC simulations of a flat plate within PICLas at angles of attack ranging from $0\,^{\circ}$ to $90\,^{\circ}$. 
Approximately $500\,000$ particle interactions are computed for every angle of attack. 
The aerodynamic coefficients are derived from the forces exerted on the plate by equations \eqref{eq:drag_coefficient} and \eqref{eq:lift_coefficient}.
\begin{equation} \label{eq:drag_coefficient}
    C_{\text{d}} = \frac{F_{\text{x}}}{\frac{1}{2} m n_{\infty} u_{\infty}^2 A}
\end{equation}
\begin{equation} \label{eq:lift_coefficient}
    C_{\ell} = \frac{F_{\text{y}}}{\frac{1}{2} m n_{\infty} u_{\infty}^2 A}
\end{equation}

\begin{figure}[tb]
    \centering
    \begin{subfigure}[t]{0.99\columnwidth}
        \centering
  \tikzsetnextfilename{cl}%
  \input{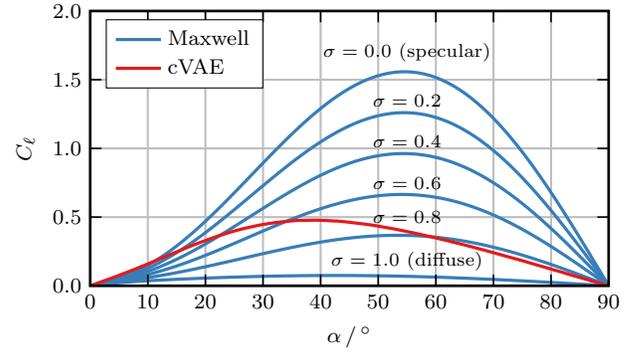}%

        \caption{Lift Coefficient}
    \end{subfigure}
    \begin{subfigure}[t]{0.99\columnwidth}
        \centering
  \tikzsetnextfilename{cd}%
  \input{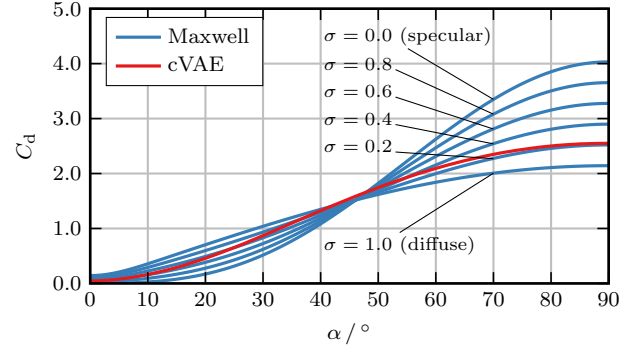}%

        \caption{Drag Coefficient}
    \end{subfigure}
    \begin{subfigure}[t]{0.99\columnwidth}
        \centering
  \tikzsetnextfilename{cl_cd_ratio}%
  \input{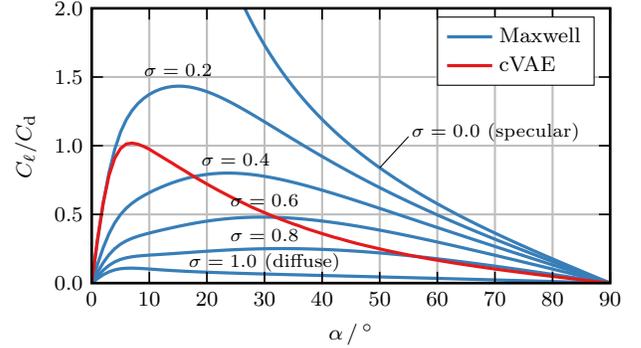}%

        \caption{Lift-to-Drag Ratio}
    \end{subfigure}
    \caption{Comparison of aerodynamic coefficients between cVAE model and Maxwell model (with $T_{\text{r}} = T_{\text{w}}$) for a flat plate in VLEO conditions.}
    \label{fig:aerodynamiccoefficients}
\end{figure}

The lift coefficient $C_{\ell}$, drag coefficient $C_{\text{d}}$, and the lift-to-drag ratio $C_{\ell}/C_{\text{d}}$ are shown in Figure \ref{fig:aerodynamiccoefficients}.
As can be seen, the Maxwell model deviates significantly from the cVAE results with regard to the shape of the curve and the absolute values.
For low angles of attack, the aerodynamic coefficients of the cVAE model are close to the Maxwell model with a high specularity.
As the angle of attack increases, the cVAE model gets closer to coefficients with high diffuse reflection.
Therefore, the cVAE can predict the change in scattering behavior from a (quasi-) specular to a diffuse one, which was recognized by the MD results in Section \ref{subsec:md_results}.
It can also be observed that the cVAE model achieves the maximum lift coefficient at lower angles of attack compared to the Maxwell model.
Moreover, the maximum lift-to-drag ratio is achieved at lower angles of attack for the cVAE model, with a rapid decline observed in the vicinity of the peak value.
Consequently, the design of an efficient geometry, for example for an aerodynamic attitude and orbit control system capable of generating higher lift than drag, has limited flexibility within this narrow operational range.

As previously mentioned, the aerodynamic coefficients of the cVAE model can be used in a panel method to improve force calculations for satellites. 
Since a satellite consists of individual surfaces rather than a single flat plate, it is necessary to provide the ram and wake surface coefficients separately instead of using the overall plate coefficients. 
These specific coefficients are presented in Table \ref{tab:aero_coeff_wake_ram} in the Appendix.

\subsection{Angular Distribution of the Molecular Flux} \label{subsec:angular_distribution}

\begin{figure}[tb]
    \centering
  \tikzsetnextfilename{sketch}%
  \input{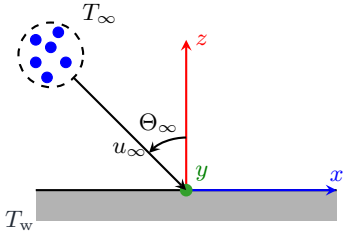}%

    \caption{Schematic representation of the DSMC simulation setup illustrating incident particles with temperature $T_{\infty}$, bulk velocity $u_\infty$, and bulk polar angle $\Theta_{\infty}$, approaching a surface of temperature $T_{\text{w}}$, used to derive the angular distribution of the reflected molecular flux.}
    \label{fig:angular_distribution_experiment_sketch}
\end{figure}

Based on the findings from the previous Section, it appears that adjusting the accommodation coefficient $\sigma$ within the Maxwell model could potentially align it with the aerodynamic coefficients of the cVAE model.
To check this hypothesis, we now examine the angular distribution of the molecular flux $\Phi$, which is defined as the number of particles passing through a unit area per unit time.
Especially, we compare the cVAE model with the Maxwell model using two different accommodation coefficient of $\sigma = 0.2$ and $\sigma = 0.8$.
This is due to the fact that the lift-to-drag ratio of the cVAE closely aligns to the Maxwell model with $\sigma = 0.2$ at low angles of attack and to the Maxwell model with $\sigma = 0.8$ at high angles of attack.
To determine the angular distribution of the molecular flux for reflected particles on a flat surface, particle interactions involving $500\,000$ particles were simulated using the DSMC method implemented in PICLas.
Two distinct bulk polar angles were analyzed: $\Theta_{\infty} = 20^\circ$ and $\Theta_{\infty} = 85^\circ$.
The schematic setup of this DSMC simulation is shown in Figure \ref{fig:angular_distribution_experiment_sketch}.
Keep in mind that the bulk polar angle is measured against the surface normal, whereas the angle of attack in the previous Section \ref{subsec:aerodynamic_coefficients} is defined relative to the bulk inflow.
Therefore, a bulk polar angles of $20^\circ$ and $85^\circ$ would correspond to $70^\circ$ and $5^\circ$ angles of attack, respectively.
The simulations of $500\,000$ particle interactions took $\approx 2.8\,\text{sec}$ with the cVAE model, while the Maxwell model is considerably faster with only $\approx 0.4\,\text{sec}$.

\begin{figure*}[tb]
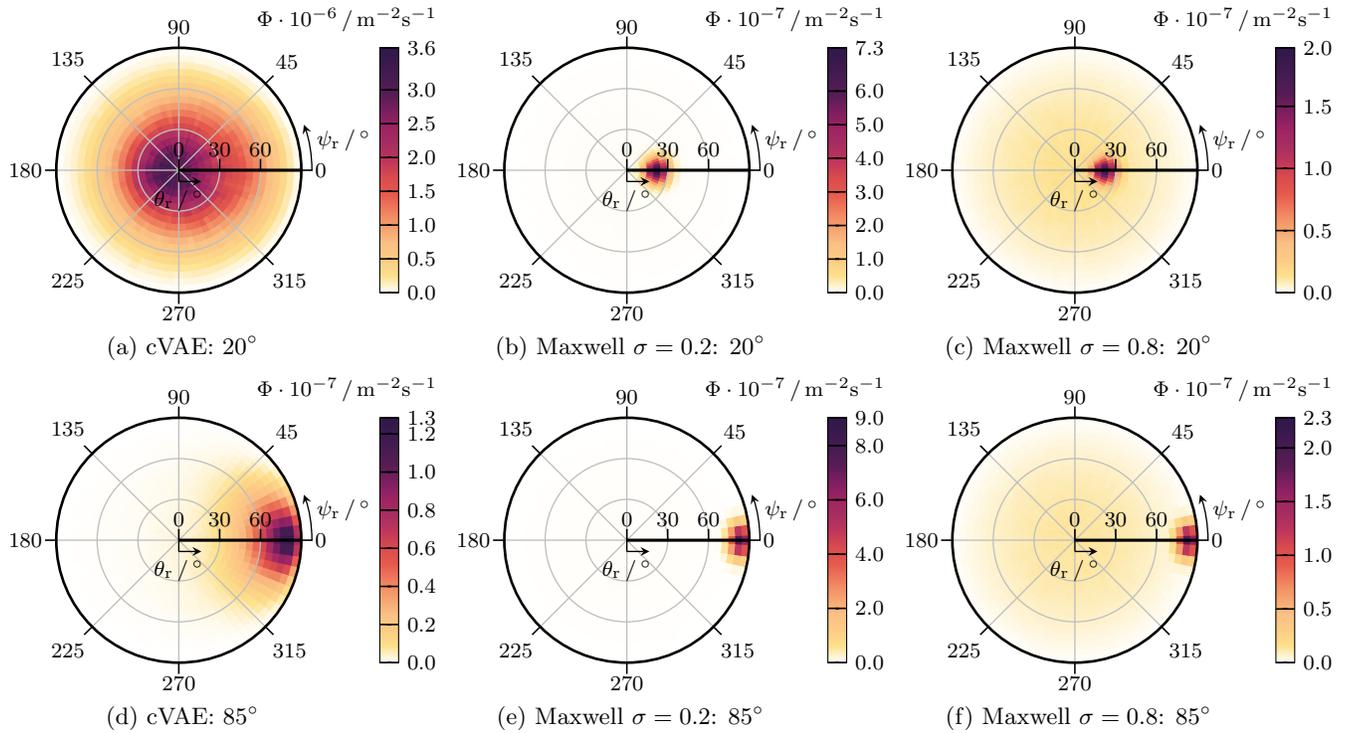

    \def\subfigwidth{0.27\textwidth}
    \centering
    \begin{subfigure}[t]{\subfigwidth}
        \centering
  \tikzsetnextfilename{angular_distribution_cvae_20}%
  \input{sections/results/application/molecular_flux/tikzpictures/20_deg/angular_distribution_cvae_20.tex}%

        \vspace{-5mm}
        \caption{cVAE: $20^\circ$}
        \label{subfig:angular_distribution_20_deg_cvae}
    \end{subfigure}
    \hspace{9mm}
    \begin{subfigure}[t]{\subfigwidth}
        \centering
  \tikzsetnextfilename{angular_distribution_maxwell_0.2_1.0_20}%
  \input{sections/results/application/molecular_flux/tikzpictures/20_deg/angular_distribution_maxwell_0.2_1.0_20.tex}%

        \vspace{-5mm}
        \caption{Maxwell $\sigma = 0.2$: $20^\circ$}
        \label{subfig:angular_distribution_20_deg_maxwell_0.2}
    \end{subfigure}
    \hspace{9mm}
    \begin{subfigure}[t]{\subfigwidth}
        \centering
  \tikzsetnextfilename{angular_distribution_maxwell_0.8_1.0_20}%
  \input{sections/results/application/molecular_flux/tikzpictures/20_deg/angular_distribution_maxwell_0.8_1.0_20.tex}%

        \vspace{-5mm}
        \caption{Maxwell $\sigma = 0.8$: $20^\circ$}
        \label{subfig:angular_distribution_20_deg_maxwell_0.8}
    \end{subfigure}
    \hspace{9mm}

    \begin{subfigure}[t]{\subfigwidth}
        \centering
  \tikzsetnextfilename{angular_distribution_cvae_85}%
  \input{sections/results/application/molecular_flux/tikzpictures/85_deg/angular_distribution_cvae_85.tex}%

        \vspace{-5mm}
        \caption{cVAE: $85^\circ$}
        \label{subfig:angular_distribution_85_deg_cvae}
    \end{subfigure}
    \hspace{9mm}
    \begin{subfigure}[t]{\subfigwidth}
        \centering
  \tikzsetnextfilename{angular_distribution_maxwell_0.2_1.0_85}%
  \input{sections/results/application/molecular_flux/tikzpictures/85_deg/angular_distribution_maxwell_0.2_1.0_85.tex}%

        \vspace{-5mm}
        \caption{Maxwell $\sigma = 0.2$: $ 85^\circ$}
        \label{subfig:angular_distribution_85_deg_maxwell_0.2}
    \end{subfigure}
    \hspace{9mm}
    \begin{subfigure}[t]{\subfigwidth}
        \centering
  \tikzsetnextfilename{angular_distribution_maxwell_0.8_1.0_85}%
  \input{sections/results/application/molecular_flux/tikzpictures/85_deg/angular_distribution_maxwell_0.8_1.0_85.tex}%

        \vspace{-5mm}
        \caption{Maxwell $\sigma = 0.8$: $85^\circ$}
        \label{subfig:angular_distribution_85_deg_maxwell_0.8}
    \end{subfigure}
    \hspace{9mm}

    \caption{Angular distribution of the molecular flux $\Phi$ for a bulk polar angle of $\Theta_{\infty} = 20^\circ$ and $\Theta_{\infty} = 85^\circ$ for the cVAE model, and Maxwell model ($\sigma = 0.2 $ and $\sigma=0.8$) under VLEO conditions.}
    \label{fig:angular_distribution}
\end{figure*}

In Figure \ref{fig:angular_distribution} the angular distribution of the molecular flux is shown for the cVAE model and the Maxwell model.
Our previously shown results of the MD simulations, as well as recent molecular beam experiments \cite{xu_hyperthermal_2025}, have shown that the angular distribution of the molecular flux depends on the incident angle.
For low polar angles, the angular distribution is dominated by diffuse reflection, while for high polar angles, the angular distribution tends to be quasi-specular.
The cVAE model is able to capture this shift in the angular distribution, as can be seen in Figures \ref{subfig:angular_distribution_20_deg_cvae} and \ref{subfig:angular_distribution_85_deg_cvae}.
Compared to the Maxwell model in Figure \ref{subfig:angular_distribution_20_deg_maxwell_0.2}/\ref{subfig:angular_distribution_20_deg_maxwell_0.8} and \ref{subfig:angular_distribution_20_deg_maxwell_0.2}/\ref{subfig:angular_distribution_85_deg_maxwell_0.8}, which is unable to predict this shift.
The reason lies in the assumption of a superposition of diffuse and specular reflections independent of the incoming angle.
Moreover, the quasi-specular reflection can never be captured by the Maxwell model.
The dependence of the angular distribution and the change in reflection behavior have a significant impact on the design of the inlet for atmospheric breathing electric propulsion systems.
Additionally, the correct molecular flux becomes increasingly significant if a satellite has large structures and the particles experience multiple reflections.

We can conclude that the variation in aerodynamic coefficients with increasing angles of attack can be attributed to the transition in scattering behavior from (quasi-) specular to diffuse. 
Although it may be possible to adjust the accommodation coefficients to approximate the aerodynamic coefficients of the cVAE model, the Maxwell model remains incapable of accurately capturing the molecular flux distribution. 

\section{Conclusion} \label{sec:conclusion}

As a result of the residual atmosphere in the VLEO regime satellites experience an atmospheric drag limiting their lifetimes.
In order to make VLEO more accessible for future missions, it is crucial to understand the GSIs in this regime for various applications, such as lifetime predictions, fuel requirements, aerodynamic optimization, and designing efficient intakes for atmospheric breathing electric propulsion systems.
The DSMC method is widely used for the aerodynamic optimization of satellites, due to the non-equilibrium flow regime.
Existing GSI models in DSMC, such as the Maxwell model, are unable to accurately predict the interactions of gas molecules with surfaces under VLEO conditions.
The Maxwell model assumes a superposition of diffuse and specular reflection controlled by a constant accommodation coefficient, which is usually unknown.
However, MD simulations performed in this study as well as molecular beam experiments in the literature have shown that GSIs are much more complex and depend on various parameters such as the incident angle and velocity magnitude.
Therefore, we introduced a novel approach for modeling GSIs in VLEO conditions using a cVAE.
MD simulations were performed to generate a dataset of GSIs, which was then used to train the cVAE model.
By embedding microscopic level insights into mesoscopic DSMC simulations, the cVAE model provides a seamless connection between microscopic interactions and macroscopic behavior.
In order to make the data generation process as efficient as possible, we selected specific incident velocities and simulated several impacts to derive the scattering kernels for the respective incident velocities.
The MD results show that the reflection behavior changes from a diffuse reflection at low incident polar angles to a (quasi-) specular reflection at high incident polar angles.
After training, the decoder of the cVAE model is able to predict the reflected velocity distribution for any given incident velocity vector, as could be shown on the validation and extrapolation data.
Therefore, the model can also be used for different gas temperatures and orbit velocities in the VLEO regime.

The results of the aerodynamic coefficients and the angular distributions of the molecular flow show significant improvements over the current most commonly used model for GSIs, the Maxwell model.
The cVAE model is able to predict the shift from diffuse reflection at low polar angles to quasi-specular reflection at high polar angles.
As a result, the general shape of the aerodynamic coefficients, the magnitudes of the coefficients, and the reflected molecular flux are significantly different from the Maxwell model.
Therefore, satellite aerodynamic optimization of satellites as well as the design of intakes for atmospheric breathing electric propulsion systems can be improved by using the cVAE model instead of the Maxwell model with guessed accommodation coefficients between $0.8$ and $1.0$.
Compared to other MD trained scattering models, our model is directly trained on MD derived scattering kernels, demonstrates reliable performance with limited data, and possesses an extrapolation capability.

Although we included the effects of atomic roughness through an amorphous surface and the influence of adsorbed atomic oxygen on the scattering behavior, the cVAE model currently neglects the influence of macroscopic roughness on a nanometer or micrometer scale. 
The roughness will probably change the scattering to more diffuse reflection.
However, future work will focus on the inclusion of these parameters in the model.

\begin{acknowledgments}
This work is funded by the Deutsche Forschungsgemeinschaft project number 516238647 - SFB1667/1 (ATLAS - Advancing Technologies for Low-Altitude Satellites). The authors also thank the High
Performance Computing Center Stuttgart (HLRS) for granting the computational time that allowed the execution of the presented molecular dynamics simulations.
\end{acknowledgments}

\section*{Data Availability Statement}
The data and code that support the findings of this study are available from the corresponding author upon reasonable request.

\section*{Conflict of Interest}
The authors declare that they have no conflicts of interest, financial or otherwise, that could have influenced the results or interpretation of the reported research.

\bibliography{library}

\appendix
\onecolumngrid

\section{Validation} \label{sec:validation}
\begin{figure*}[ht]
    \centering
    \begin{minipage}{0.99\textwidth}
        \centering
        \begin{minipage}{0.324\textwidth}
            \centering
  \tikzsetnextfilename{v_mag_7923.6_theta_0.0_t1_directions}%
  \input{sections/appendix/tikzpictures/histogram_v_mag_7923.6_theta_0.0/v_mag_7923.6_theta_0.0_t1_directions.tex}%

        \end{minipage}
        \begin{minipage}{0.324\textwidth}
            \centering
  \tikzsetnextfilename{v_mag_7923.6_theta_0.0_t2_directions}%
  \input{sections/appendix/tikzpictures/histogram_v_mag_7923.6_theta_0.0/v_mag_7923.6_theta_0.0_t2_directions.tex}%

        \end{minipage}
        \begin{minipage}{0.324\textwidth}
            \centering
  \tikzsetnextfilename{v_mag_7923.6_theta_0.0_n_directions}%
  \input{sections/appendix/tikzpictures/histogram_v_mag_7923.6_theta_0.0/v_mag_7923.6_theta_0.0_n_directions.tex}%

        \end{minipage}
        \vspace{-2mm}
        \subcaption{$|\bm{v}_{\text{i}}| = 7923.6 \, \text{m}/\text{s}$, $\theta_{\text{i}} = 0.0 \, ^\circ$}
    \end{minipage}
    \begin{minipage}{0.99\textwidth}
        \centering
        \begin{minipage}{0.324\textwidth}
            \centering
  \tikzsetnextfilename{v_mag_7923.6_theta_10.0_t1_directions}%
  \input{sections/appendix/tikzpictures/histogram_v_mag_7923.6_theta_10.0/v_mag_7923.6_theta_10.0_t1_directions.tex}%

        \end{minipage}
        \begin{minipage}{0.324\textwidth}
            \centering
  \tikzsetnextfilename{v_mag_7923.6_theta_10.0_t2_directions}%
  \input{sections/appendix/tikzpictures/histogram_v_mag_7923.6_theta_10.0/v_mag_7923.6_theta_10.0_t2_directions.tex}%

        \end{minipage}
        \begin{minipage}{0.324\textwidth}
            \centering
  \tikzsetnextfilename{v_mag_7923.6_theta_10.0_n_directions}%
  \input{sections/appendix/tikzpictures/histogram_v_mag_7923.6_theta_10.0/v_mag_7923.6_theta_10.0_n_directions.tex}%

        \end{minipage}
        \vspace{-2mm}
        \subcaption{$|\bm{v}_{\text{i}}| = 7923.6 \, \text{m}/\text{s}$, $\theta_{\text{i}} = 10.0 \, ^\circ$}
    \end{minipage}
    \begin{minipage}{0.99\textwidth}
        \centering
        \begin{minipage}{0.324\textwidth}
            \centering
  \tikzsetnextfilename{v_mag_7923.6_theta_20.0_t1_directions}%
  \input{sections/appendix/tikzpictures/histogram_v_mag_7923.6_theta_20.0/v_mag_7923.6_theta_20.0_t1_directions.tex}%

        \end{minipage}
        \begin{minipage}{0.324\textwidth}
            \centering
  \tikzsetnextfilename{v_mag_7923.6_theta_20.0_t2_directions}%
  \input{sections/appendix/tikzpictures/histogram_v_mag_7923.6_theta_20.0/v_mag_7923.6_theta_20.0_t2_directions.tex}%

        \end{minipage}
        \begin{minipage}{0.324\textwidth}
            \centering
  \tikzsetnextfilename{v_mag_7923.6_theta_20.0_n_directions}%
  \input{sections/appendix/tikzpictures/histogram_v_mag_7923.6_theta_20.0/v_mag_7923.6_theta_20.0_n_directions.tex}%

        \end{minipage}
        \vspace{-2mm}
        \subcaption{$|\bm{v}_{\text{i}}| = 7923.6 \, \text{m}/\text{s}$, $\theta_{\text{i}} = 20.0 \, ^\circ$}
    \end{minipage}
    \begin{minipage}{0.99\textwidth}
        \centering
        \begin{minipage}{0.324\textwidth}
            \centering
  \tikzsetnextfilename{v_mag_7923.6_theta_30.0_t1_directions}%
  \input{sections/appendix/tikzpictures/histogram_v_mag_7923.6_theta_30.0/v_mag_7923.6_theta_30.0_t1_directions.tex}%

        \end{minipage}
        \begin{minipage}{0.324\textwidth}
            \centering
  \tikzsetnextfilename{v_mag_7923.6_theta_30.0_t2_directions}%
  \input{sections/appendix/tikzpictures/histogram_v_mag_7923.6_theta_30.0/v_mag_7923.6_theta_30.0_t2_directions.tex}%

        \end{minipage}
        \begin{minipage}{0.324\textwidth}
            \centering
  \tikzsetnextfilename{v_mag_7923.6_theta_30.0_n_directions}%
  \input{sections/appendix/tikzpictures/histogram_v_mag_7923.6_theta_30.0/v_mag_7923.6_theta_30.0_n_directions.tex}%

        \end{minipage}
        \vspace{-2mm}
        \subcaption{$|\bm{v}_{\text{i}}| = 7923.6 \, \text{m}/\text{s}$, $\theta_{\text{i}} = 30.0 \, ^\circ$}
    \end{minipage}
    \begin{minipage}{0.99\textwidth}
        \centering
        \begin{minipage}{0.324\textwidth}
            \centering
  \tikzsetnextfilename{v_mag_7923.6_theta_40.0_t1_directions}%
  \input{sections/appendix/tikzpictures/histogram_v_mag_7923.6_theta_40.0/v_mag_7923.6_theta_40.0_t1_directions.tex}%

        \end{minipage}
        \begin{minipage}{0.324\textwidth}
            \centering
  \tikzsetnextfilename{v_mag_7923.6_theta_40.0_t2_directions}%
  \input{sections/appendix/tikzpictures/histogram_v_mag_7923.6_theta_40.0/v_mag_7923.6_theta_40.0_t2_directions.tex}%

        \end{minipage}
        \begin{minipage}{0.324\textwidth}
            \centering
  \tikzsetnextfilename{v_mag_7923.6_theta_40.0_n_directions}%
  \input{sections/appendix/tikzpictures/histogram_v_mag_7923.6_theta_40.0/v_mag_7923.6_theta_40.0_n_directions.tex}%

        \end{minipage}
        \vspace{-2mm}
        \subcaption{$|\bm{v}_{\text{i}}| = 7923.6 \, \text{m}/\text{s}$, $\theta_{\text{i}} = 40.0 \, ^\circ$}
    \end{minipage}
    \vspace{-2mm}
    \caption{Probability density functions of the conditional Variational Autoencoder (\textcolor{customred}{\rule[0.5ex]{0.5cm}{1pt}}) and the Molecular Dynamics simulation (\textcolor{customblue}{\rule[0.5ex]{0.5cm}{1pt}}) for the velocity magnitude $|\bm{v}_{\text{i}}| = 7923.6 \, \text{m}/\text{s}$ and angles $\theta_{\text{i}}$ from $0.0\,^{\circ}$ to $40.0\,^{\circ}$.}
    \label{fig:histograms_md_cVAE_0_40}
\end{figure*}

\begin{figure*}[ht]
    \centering
    \begin{minipage}{0.99\textwidth}
        \centering
        \begin{minipage}{0.324\textwidth}
            \centering
  \tikzsetnextfilename{v_mag_7923.6_theta_50.0_t1_directions}%
  \input{sections/appendix/tikzpictures/histogram_v_mag_7923.6_theta_50.0/v_mag_7923.6_theta_50.0_t1_directions.tex}%

        \end{minipage}
        \begin{minipage}{0.324\textwidth}
            \centering
  \tikzsetnextfilename{v_mag_7923.6_theta_50.0_t2_directions}%
  \input{sections/appendix/tikzpictures/histogram_v_mag_7923.6_theta_50.0/v_mag_7923.6_theta_50.0_t2_directions.tex}%

        \end{minipage}
        \begin{minipage}{0.324\textwidth}
            \centering
  \tikzsetnextfilename{v_mag_7923.6_theta_50.0_n_directions}%
  \input{sections/appendix/tikzpictures/histogram_v_mag_7923.6_theta_50.0/v_mag_7923.6_theta_50.0_n_directions.tex}%

        \end{minipage}
        \vspace{-2mm}
        \subcaption{$|\bm{v}_{\text{i}}| = 7923.6 \, \text{m}/\text{s}$, $\theta_{\text{i}} = 50.0 \, ^\circ$}
    \end{minipage}
    \begin{minipage}{0.99\textwidth}
        \centering
        \begin{minipage}{0.324\textwidth}
            \centering
  \tikzsetnextfilename{v_mag_7923.6_theta_60.0_t1_directions}%
  \input{sections/appendix/tikzpictures/histogram_v_mag_7923.6_theta_60.0/v_mag_7923.6_theta_60.0_t1_directions.tex}%

        \end{minipage}
        \begin{minipage}{0.324\textwidth}
            \centering
  \tikzsetnextfilename{v_mag_7923.6_theta_60.0_t2_directions}%
  \input{sections/appendix/tikzpictures/histogram_v_mag_7923.6_theta_60.0/v_mag_7923.6_theta_60.0_t2_directions.tex}%

        \end{minipage}
        \begin{minipage}{0.324\textwidth}
            \centering
  \tikzsetnextfilename{v_mag_7923.6_theta_60.0_n_directions}%
  \input{sections/appendix/tikzpictures/histogram_v_mag_7923.6_theta_60.0/v_mag_7923.6_theta_60.0_n_directions.tex}%

        \end{minipage}
        \vspace{-2mm}
        \subcaption{$|\bm{v}_{\text{i}}| = 7923.6 \, \text{m}/\text{s}$, $\theta_{\text{i}} = 60.0 \, ^\circ$}
    \end{minipage}
    \begin{minipage}{0.99\textwidth}
        \centering
        \begin{minipage}{0.324\textwidth}
            \centering
  \tikzsetnextfilename{v_mag_7923.6_theta_70.0_t1_directions}%
  \input{sections/appendix/tikzpictures/histogram_v_mag_7923.6_theta_70.0/v_mag_7923.6_theta_70.0_t1_directions.tex}%

        \end{minipage}
        \begin{minipage}{0.324\textwidth}
            \centering
  \tikzsetnextfilename{v_mag_7923.6_theta_70.0_t2_directions}%
  \input{sections/appendix/tikzpictures/histogram_v_mag_7923.6_theta_70.0/v_mag_7923.6_theta_70.0_t2_directions.tex}%

        \end{minipage}
        \begin{minipage}{0.324\textwidth}
            \centering
  \tikzsetnextfilename{v_mag_7923.6_theta_70.0_n_directions}%
  \input{sections/appendix/tikzpictures/histogram_v_mag_7923.6_theta_70.0/v_mag_7923.6_theta_70.0_n_directions.tex}%

        \end{minipage}
        \vspace{-2mm}
        \subcaption{$|\bm{v}_{\text{i}}| = 7923.6 \, \text{m}/\text{s}$, $\theta_{\text{i}} = 70.0 \, ^\circ$}
    \end{minipage}
    \begin{minipage}{0.99\textwidth}
        \centering
        \begin{minipage}{0.324\textwidth}
            \centering
  \tikzsetnextfilename{v_mag_7923.6_theta_80.0_t1_directions}%
  \input{sections/appendix/tikzpictures/histogram_v_mag_7923.6_theta_80.0/v_mag_7923.6_theta_80.0_t1_directions.tex}%

        \end{minipage}
        \begin{minipage}{0.324\textwidth}
            \centering
  \tikzsetnextfilename{v_mag_7923.6_theta_80.0_t2_directions}%
  \input{sections/appendix/tikzpictures/histogram_v_mag_7923.6_theta_80.0/v_mag_7923.6_theta_80.0_t2_directions.tex}%

        \end{minipage}
        \begin{minipage}{0.324\textwidth}
            \centering
  \tikzsetnextfilename{v_mag_7923.6_theta_80.0_n_directions}%
  \input{sections/appendix/tikzpictures/histogram_v_mag_7923.6_theta_80.0/v_mag_7923.6_theta_80.0_n_directions.tex}%

        \end{minipage}
        \vspace{-2mm}
        \subcaption{$|\bm{v}_{\text{i}}| = 7923.6 \, \text{m}/\text{s}$, $\theta_{\text{i}} = 80.0 \, ^\circ$}
    \end{minipage}
    \vspace{-2mm}
    \caption{Probability density functions of the conditional Variational Autoencoder (\textcolor{customred}{\rule[0.5ex]{0.5cm}{1pt}}) and the Molecular Dynamics simulation (\textcolor{customblue}{\rule[0.5ex]{0.5cm}{1pt}}) for the velocity magnitude $|\bm{v}_{\text{i}}| = 7923.6 \, \text{m}/\text{s}$ and angles $\theta_{\text{i}}$ from $50.0\,^{\circ}$ to $80.0\,^{\circ}$.}
    \label{fig:histograms_md_cVAE_50_80}
\end{figure*}

\begin{samepage}
    \section{Aerodynamic coefficients}
    \begin{table}[H]
        \setlength{\tabcolsep}{5pt}
        \centering
        \caption{Aerodynamic coefficients of the ram and wake surfaces of a flat plate determined with the cVAE model under VLEO conditions, as specified in Table \ref{tab:env_condition}.}
        \label{tab:aero_coeff_wake_ram}
        \begin{minipage}[t]{0.48\textwidth}
            \centering
            \begin{filecontents*}{sections/appendix/aerodynamic_coefficients_panel_method_0_45.csv}
AoA,Clram,Cdram,Clwake,Cdwake
0.0,0.029375455865880053,0.022969020659585695,-0.029378384156817424,0.02297629623491647
1.0,0.03744653397916193,0.030077008031983658,-0.022581769463367127,0.01715605130997958
2.0,0.04676451963863769,0.03856305167352887,-0.016968151470336243,0.01254582062141745
3.0,0.057404341130666836,0.04854929568601092,-0.01244664187162758,0.00895691409660611
4.0,0.06919266608068343,0.06014808440705376,-0.008919427703942192,0.0062266826330858146
5.0,0.08212251266258373,0.07314780800754385,-0.006233847764910955,0.004238236135911338
6.0,0.09614463027253599,0.08799688892700162,-0.004233793180994701,0.0027868480792328535
7.0,0.11090224676524468,0.10443571628556508,-0.0028118872726789956,0.001797023878546151
8.0,0.1266385324072098,0.12234389079157454,-0.0018131281021085317,0.0011249584555471969
9.0,0.14282916729064876,0.1416787004846276,-0.0011376902756256276,0.0006830145502935161
10.0,0.1598068374811365,0.16317476622662366,-0.0006940228943824426,0.0004035930401212204
11.0,0.17689225496024982,0.18577634390808562,-0.00041087516428886485,0.00023138091771309768
12.0,0.194230986648961,0.21003544017403059,-0.00023678305335736232,0.00012899343045711905
13.0,0.21179290199999745,0.2355906347746857,-0.0001326922818745899,6.969612850515849e-5
14.0,0.22943713713075153,0.2631233086802303,-7.226558206913866e-5,3.6692018672433276e-5
15.0,0.24705977627722395,0.29145141442678263,-3.815481834395366e-5,1.8635298055979224e-5
16.0,0.26424833265906156,0.3221206684262991,-1.9635194756026604e-5,9.260921199680112e-6
17.0,0.28131952391057174,0.3539100361048616,-9.80465668339313e-6,4.444262475760161e-6
18.0,0.2980000739628811,0.3869312618015575,-4.7756256917403125e-6,2.0826096995539247e-6
19.0,0.31387865573277646,0.42157154099890026,-2.2613951882816338e-6,9.471573877076285e-7
20.0,0.32972193656589444,0.45710852686806114,-1.0434471245759262e-6,4.1983485294064734e-7
21.0,0.34466006123065485,0.49469238786226427,-4.6938002197060463e-7,1.808917290768512e-7
22.0,0.35949355646569164,0.5321869142247728,-2.0602634414549633e-7,7.58676093812265e-8
23.0,0.37308296669454055,0.5720090084339461,-1.2268669087598713e-7,-1.144070030569634e-8
24.0,0.38625954050698863,0.6116592499656929,-0.0,-0.0
25.0,0.3982815340589537,0.6533746424270191,-0.0,-0.0
26.0,0.40931279778099583,0.6957754658258896,-0.0,-0.0
27.0,0.42005343085786284,0.7379043331127854,-0.0,-0.0
28.0,0.4296496616737772,0.7813573021433402,-0.0,-0.0
29.0,0.43822100969920064,0.8258663198536258,-0.0,-0.0
30.0,0.44607073527539004,0.8707098142067964,-0.0,-0.0
31.0,0.45244453812954266,0.9151865758397787,-0.0,-0.0
32.0,0.45934972264471735,0.9601913358846619,-0.0,-0.0
33.0,0.4643002768133419,1.005679708457244,-0.0,-0.0
34.0,0.46822726094239364,1.0513961128618257,-0.0,-0.0
35.0,0.4722873880740148,1.0964905046585396,-0.0,-0.0
36.0,0.4746389608847326,1.142143098088083,-0.0,-0.0
37.0,0.4760162135558514,1.1875472626936703,-0.0,-0.0
38.0,0.47689006475992307,1.233718882263641,-0.0,-0.0
39.0,0.47769102097792093,1.277404819942964,-0.0,-0.0
40.0,0.4762267010072548,1.3235372269954802,-0.0,-0.0
41.0,0.47479871877685054,1.3672539686876628,-0.0,-0.0
42.0,0.4732989982656917,1.411316747918407,-0.0,-0.0
43.0,0.4701640718337886,1.4551932394155371,-0.0,-0.0
44.0,0.4669765781967516,1.497500409160734,-0.0,-0.0
45.0,0.4631027103232773,1.5409327076165784,-0.0,-0.0
\end{filecontents*}
\pgfplotstabletypeset[
                col sep=comma,
                header=true,
                every head row/.style={before row=\hline\hline, after row=\hline},
                every last row/.style={after row=\hline\hline},
                columns/AoA/.style={
                    column name={$\alpha\,/\,^{\circ}$},
                    column type={>{\centering\arraybackslash}m{0.8cm}<{\rule{0pt}{0.3cm}}}
                },
                columns/Clram/.style={
                    column name={$C_{\ell,\text{ram}}$},
                    column type={>{\centering\arraybackslash}m{1.5cm}<{\rule{0pt}{0.3cm}}},
                    fixed, precision=3, zerofill
                },
                columns/Cdram/.style={
                    column name={$C_{\text{d},\text{ram}}$},
                    column type={>{\centering\arraybackslash}m{1.5cm}<{\rule{0pt}{0.3cm}}},
                    fixed, precision=3, zerofill
                },
                columns/Clwake/.style={
                    column name={$C_{\ell,\text{wake}}$},
                    column type={>{\centering\arraybackslash}m{1.5cm}<{\rule{0pt}{0.3cm}}},
                    fixed, precision=3, zerofill
                },
                columns/Cdwake/.style={
                    column name={$C_{\text{d},\text{wake}}$},
                    column type={>{\centering\arraybackslash}m{1.5cm}<{\rule{0pt}{0.3cm}}},
                    fixed, precision=3, zerofill
                },
            ]{sections/appendix/aerodynamic_coefficients_panel_method_0_45.csv}
        \end{minipage}
        \hfill
        \begin{minipage}[t]{0.48\textwidth}
            \centering
            \begin{filecontents*}{sections/appendix/aerodynamic_coefficients_panel_method_45_90.csv}
AoA,Clram,Cdram,Clwake,Cdwake
45.0,0.4631027103232773,1.5409327076165784,-0.0,-0.0
46.0,0.4571652020311126,1.5828311439483964,-0.0,-0.0
47.0,0.45264929906200907,1.6242744287603463,-0.0,-0.0
48.0,0.44739112540185616,1.6651203645802328,-0.0,-0.0
49.0,0.43958599471038956,1.705667057431269,-0.0,-0.0
50.0,0.4342712095802754,1.7446488116052525,-0.0,-0.0
51.0,0.4266670847539333,1.7831948808926865,-0.0,-0.0
52.0,0.41898786636135243,1.821123833476282,-0.0,-0.0
53.0,0.4112912802852413,1.8578581950664466,-0.0,-0.0
54.0,0.4036235031640186,1.8941443562051157,-0.0,-0.0
55.0,0.394665054073772,1.9296155669195978,-0.0,-0.0
56.0,0.38568253276254244,1.9645315668519636,-0.0,-0.0
57.0,0.3771976749704962,1.9979623703825984,-0.0,-0.0
58.0,0.3671189655975356,2.0304659082851684,-0.0,-0.0
59.0,0.3576242177938485,2.0627246828798804,-0.0,-0.0
60.0,0.34854716752414217,2.0939751932232724,-0.0,-0.0
61.0,0.33900099681128015,2.123125150767951,-0.0,-0.0
62.0,0.328416874155361,2.1524037377852903,-0.0,-0.0
63.0,0.31776712173774213,2.180380285589894,-0.0,-0.0
64.0,0.30695318211286526,2.2071318911752145,-0.0,-0.0
65.0,0.29653761969813747,2.2330139408016856,-0.0,-0.0
66.0,0.28548088047760317,2.258099333648644,-0.0,-0.0
67.0,0.2746437381428266,2.2825690133729113,-0.0,-0.0
68.0,0.2642057457470637,2.3046216917632223,-0.0,-0.0
69.0,0.25194845822957906,2.328122664586785,-0.0,-0.0
70.0,0.24141085903729675,2.3491535027947688,-0.0,-0.0
71.0,0.22870598882269558,2.368995035598637,-0.0,-0.0
72.0,0.21712151469569296,2.3880087816989475,-0.0,-0.0
73.0,0.2053366086865225,2.405808341766528,-0.0,-0.0
74.0,0.19480680323406824,2.422771310394834,-0.0,-0.0
75.0,0.1817632447055411,2.4382329627474544,-0.0,-0.0
76.0,0.17069865280410315,2.453005625057659,-0.0,-0.0
77.0,0.15771177232683145,2.467008910963511,-0.0,-0.0
78.0,0.14574927243496177,2.4797709558785943,-0.0,-0.0
79.0,0.13371393972199438,2.4904387852679815,-0.0,-0.0
80.0,0.12126082732816575,2.5015214322331425,-0.0,-0.0
81.0,0.10956480376885316,2.509852527436894,-0.0,-0.0
82.0,0.09740787903682896,2.5193332065676555,-0.0,-0.0
83.0,0.0852682985929166,2.5265525460864113,-0.0,-0.0
84.0,0.07252338666147275,2.5325016593663263,-0.0,-0.0
85.0,0.060360198532591204,2.53864904568156,-0.0,-0.0
86.0,0.04808829091053361,2.5422021897132816,-0.0,-0.0
87.0,0.03698395562894446,2.5456218425512893,-0.0,-0.0
88.0,0.023019002403652,2.548235344956045,-0.0,-0.0
89.0,0.011557425257186829,2.5499085786626408,-0.0,-0.0
90.0,-0.00043184635582051894,2.549771592280347,-0.0,-0.0
\end{filecontents*}
\pgfplotstabletypeset[
                col sep=comma,
                header=true,
                every head row/.style={before row=\hline\hline, after row=\hline},
                every last row/.style={after row=\hline\hline},
                columns/AoA/.style={
                    column name={$\alpha\,/\,^{\circ}$},
                    column type={>{\centering\arraybackslash}m{0.8cm}<{\rule{0pt}{0.3cm}}}
                },
                columns/Clram/.style={
                    column name={$C_{\ell,\text{ram}}$},
                    column type={>{\centering\arraybackslash}m{1.5cm}<{\rule{0pt}{0.3cm}}},
                    fixed, precision=3, zerofill
                },
                columns/Cdram/.style={
                    column name={$C_{\text{d},\text{ram}}$},
                    column type={>{\centering\arraybackslash}m{1.5cm}<{\rule{0pt}{0.3cm}}},
                    fixed, precision=3, zerofill
                },
                columns/Clwake/.style={
                    column name={$C_{\ell,\text{wake}}$},
                    column type={>{\centering\arraybackslash}m{1.5cm}<{\rule{0pt}{0.3cm}}},
                    fixed, precision=3, zerofill
                },
                columns/Cdwake/.style={
                    column name={$C_{\text{d},\text{wake}}$},
                    column type={>{\centering\arraybackslash}m{1.5cm}<{\rule{0pt}{0.3cm}}},
                    fixed, precision=3, zerofill
                },
            ]{sections/appendix/aerodynamic_coefficients_panel_method_45_90.csv}
        \end{minipage}
    \end{table}
\end{samepage}

\end{document}